\documentclass[prb,aps,twocolumn,superscriptaddress,showpacs,10pt,floatfix]{revtex4-1}
\usepackage{amsmath}
\usepackage{amsfonts}
\usepackage{amssymb}
\usepackage{bm}
\usepackage{graphicx}
\usepackage{datetime}
\usepackage{color}
\usepackage{subfigure}

\begin{document}
\newcommand{\figwidth}{0.95\columnwidth}
\newcommand{\ffigwidth}{0.4\columnwidth}
\renewcommand\Re{\operatorname{\mathfrak{Re}}}
\renewcommand\Im{\operatorname{\mathfrak{Im}}}
\newcommand{\bra}[1]{\langle #1|}
\newcommand{\ket}[1]{|#1\rangle}
\newcommand{\braket}[2]{\langle #1|#2\rangle}
\newcommand{\ketbra}[2]{| #1 \rangle \langle #2|}
\newcommand{\revision}[1]{\textcolor{red}{{#1}}}
\newcommand{\warwick}{Department of Physics, University of Warwick, Coventry, CV4 7AL, United Kingdom}
\newcommand{\warwickcsc}{Centre for Scientific Computing, University of Warwick, Coventry, CV4 7AL, United Kingdom}
\newcommand{\mainz}{Dipartimento di Fisica, Universit\'a di Roma ``La Sapienza'', I-00185, Roma, Italy and Institut f\"{u}r Physik, Universit\"{a}t Mainz, D-55099 Mainz, Germany}
\title{Localization-delocalization transition for disordered cubic harmonic lattices}
\author{Sebastian D. Pinski}\affiliation{\warwick}\affiliation{\warwickcsc}
\email[Corresponding author:]{s.d.pinski@warwick.ac.uk}
\author{Walter Schirmacher}\affiliation{\mainz}
\author{Terry Whall}\affiliation{\warwick}
\author{Rudolf A. R\"omer}\affiliation{\warwick} \affiliation{\warwickcsc}
\date{$Revision: 1.104 $, compiled \today, \currenttime}
%
\begin{abstract}
We study numerically the disorder-induced localization-delocalization phase transitions that occur for mass and spring constant disorder in a three-dimensional cubic lattice with harmonic couplings. We show that, while the phase diagrams exhibit regions of stable and unstable waves, the universality of the transitions is the same for mass and spring constant disorder throughout all the phase boundaries. The combined value for the critical exponent of the localization lengths of $\nu =1.550^{+0.020}_{-0.017}$ confirms the agreement with the universality class of the standard electronic Anderson model of localization. We further support our investigation with studies of the density of states, the participation numbers and wave function statistics. 
\end{abstract}
\maketitle
%
\section{Introduction}
\label{sec-intro}
%
The disorder-induced metal-insulator transition (MIT) and the concept of Anderson localization\cite{And58} have been studied extensively for over $50$ years. Most of the attention was focused on electronic systems and their transport properties\cite{LeeR85,KraM93,BelK94,EveM08} --- indeed the acronym MIT itself suggests this. However, localization physics is of course much broader than just electrons in solid state devices and encompasses the whole realm of waves --- quantum and classical --- and their interference due to random scattering events. Recently, the interest in localization has been rekindled by its beautiful realization in cold atom systems. \cite{BilJZB08,RoaDFF08} Similarly, localization of classical waves has received new impetus from spatially resolved studies in elastic, vibrational systems.\cite{FaeSPL09} 

Theoretical work on the localization properties of harmonic solids has received somewhat less attention over the years. In our opinion, this could be due to (i) a general expectation that the vibrational problem only mimics the electronic one and (ii) the one clear feature when this is not the case --- the so-called ``boson peak" (BP)\cite{SchDG98,KanRB01} --- up to this date remains to be understood fully. In a recent paper,\cite{PinSR12} we have shown that expectation (i) is only partially true: the phase diagrams, even for just a simple cubic harmonic lattice of masses and springs, exhibit several intriguing features for both the purely mass and the purely spring constant disordered cases. A similarly distinguishing characteristic of vibrational localization is the fact that the zero frequency, i.e.\ $\omega=0$ mode that corresponds to global translational invariance, cannot be localized regardless of the amount of disorder.\cite{Rus02} 
The aforementioned BP corresponds to the appearance of a low-frequency enhancement of the density of states $g(\omega)$ with respect to Debye's $g(\omega)\propto \omega^2$ law.\cite{SchDG98,KanRB01} Most previous investigations of the localization properties of disordered vibrational modes agree that the modes near and above the BP are extended,\cite{SchDG98,SchD99,FelKAW93} i.e.\ $\omega_{\rm BP}\ll\omega_{\rm c}$, where $\omega_{\rm BP}$ denotes the BP frequency (peak of $g(\omega)/\omega^2$) and $\omega_{\rm c}$ the boundary between extended and localized states. It has been argued before via eigenvalue statistics that the states with $\omega_{\rm BP}<\omega<\omega_{\rm c}$ are governed by random-matrix statistics of the Gaussian orthogonal ensemble (GOE).\cite{SchDG98,SarMP04}
%

\begin{figure*}[tbh]
(a)\includegraphics[width=0.45\textwidth]{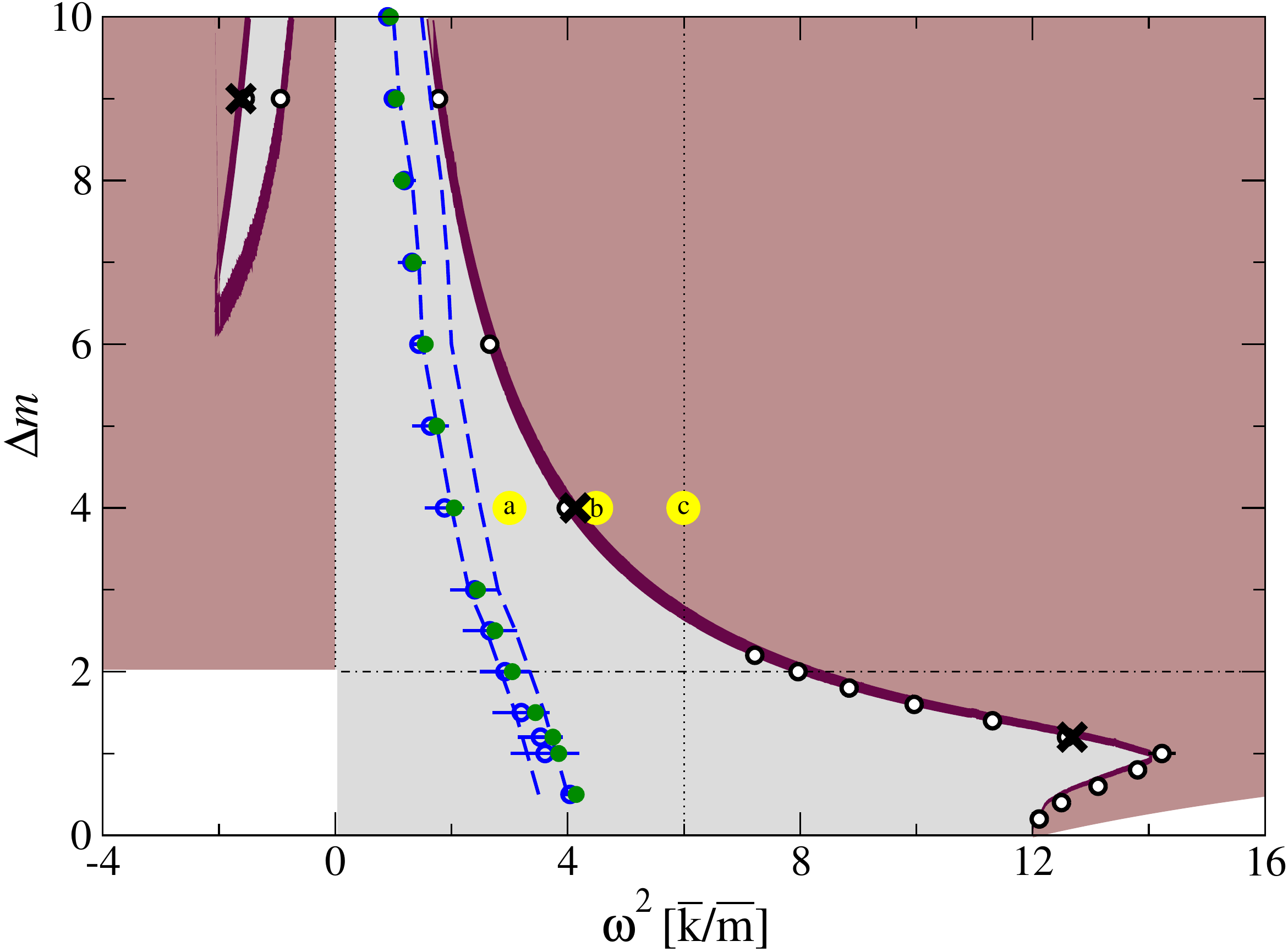}
(b)\includegraphics[width=0.45\textwidth]{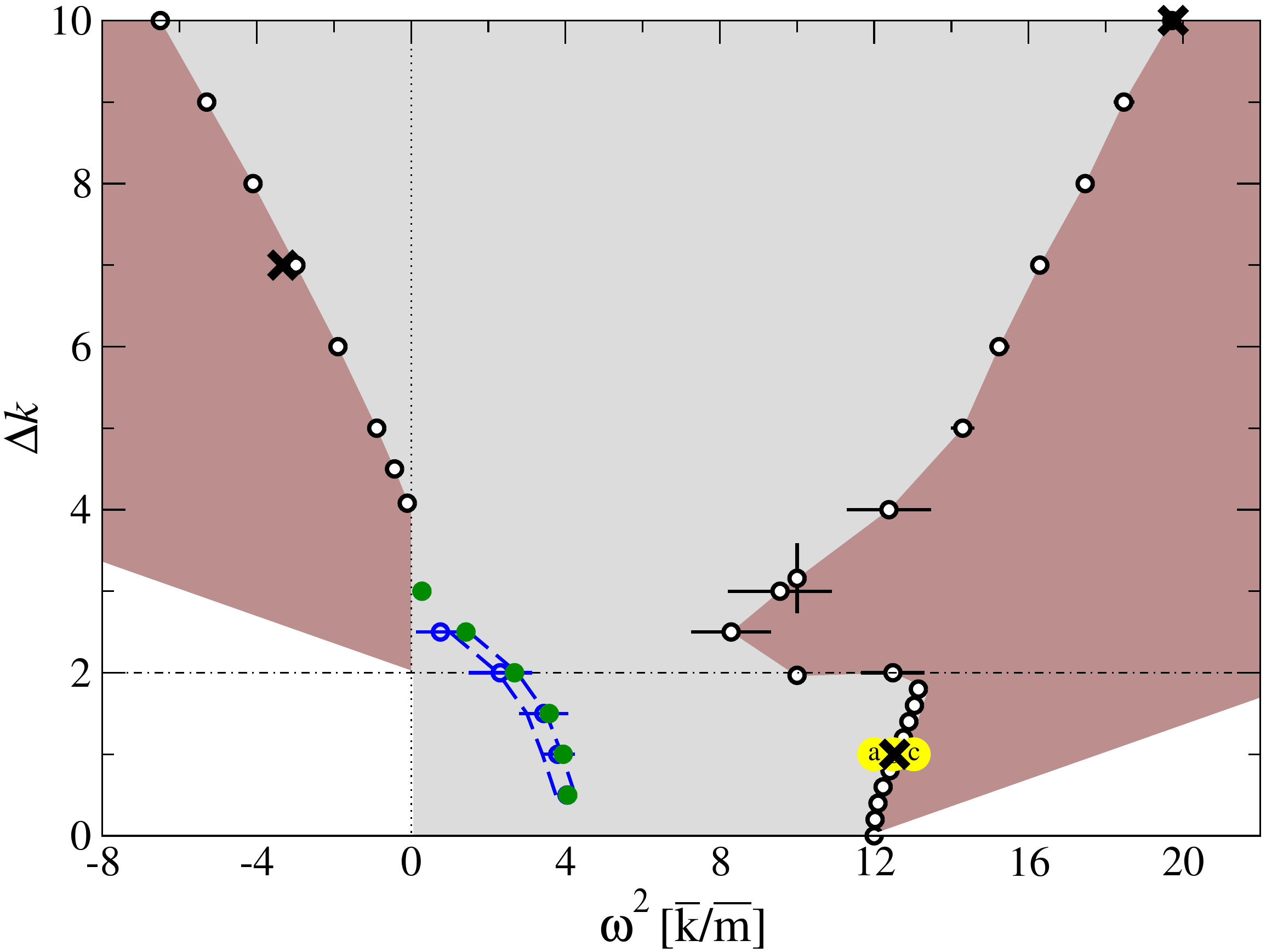}
\caption{(Color online) Phase diagrams for (a) mass ($\Delta m$ vs. $\omega^2$) and (b) spring constant ($\Delta k$ vs. $\omega^2$) disorder. Grey and (dark) brown shadings are the regions of extended and localized states, respectively. The white regions are inaccessible and their perimeter denote the band edges. The horizontal dash-dotted line indicates the border between stable and unstable regions, the dotted lines denote $\omega^2=0$ and $6$. Thick black crosses, white circles, open blue circles and closed green circles indicate (i) the three high-precision transition values obtained from FSS, (ii) the other lower-accuracy transition points, (iii) the reduced VDOS peak locations from numerics and (iv) the reduced VDOS peak locations computed via CPA, respectively, in both (a) and (b). The yellow labels \textcircled{a}--\textcircled{c} show the positions of the states in Figs.\ \ref{fig-Mstates}(a)--(c) and Figs.\ \ref{fig-Kstates}(a)--(c). The dashed blue lines indicate the range of deviations Eq.\ \eqref{eq-diffGOE} from universal wave function statistics, cp.\ also the insets of Figs.\ \ref{fig-PTS} and \ref{fig-PTS-k}. Additionally in (a) the (very dark) marroon shading denotes the critical region obtained from the transformation of the electronic phase diagram.\cite{BulSK87,PinSR12}}
\label{fig-phase}
\end{figure*}

In this paper, we present a detailed study of the vibrational localization and transport properties throughout the previously obtained phase diagrams of a cubic harmonic lattice system with either random mass or random spring constant disorder. Using large matrix diagonalization techniques, we investigate the behaviour of the vibrational density of states (VDOS) as well as the participation numbers and wave function statistics of the vibrational eigenstates. This complements earlier studies of participation ratios,\cite{CanV85,LudSTE01} level-spacing statistics\cite{SchDG98,ShiNN07} and multifractal properties.\cite{LudTE03} In particular we demonstrate that the disorder-effected states below $\omega_{\rm c}$ exhibit a modified Porter-Thomas statistics of the wave functions, which is close to the one from the GOE ensemble. In addition, we present results from a high-precision transfer matrix method (TMM) and a finite-size scaling (FSS) analysis which allow us to corroborate the phase diagrams and calculate the universality class of the mobility edges across all of the phase diagram. 
Our results have relevance in the related problem of instantaneous normal modes in glasses and supercooled liquids\cite{BemL95,BemL96,HuaW10,HuaW09} as well as acoustic metamaterials.\cite{LiuZMZ00,Hir04,ChaLF06,ZhaYF09,DinLQS07,WriC09} Here we just note that in both these classes of materials, there exist excitations which can be related to the existence of states in what is formally part of the temporally decaying, negative $\omega^2$ region of the phase diagrams shown in Fig.\ \ref{fig-phase}.

\section{Scalar model of lattice vibrations}
\label{sec-numapp}

\subsection{The clean case}
\label{sec-clean}

We shall consider masses arranged on a simple cubic lattice and connected by harmonic forces. With $\vec{u}_{j}$ denoting the deviation from the lattice equilibrium position $\vec{r}_j=(x,y,z)_j$ of a certain mass $m_{j}$ at given $x$, $y$ and $z$ lattice coordinates, we can write the classical equations of motion as
\begin{equation}
 m_{j} \ddot{\vec{u}}_{j} =
 -\sum_{\text{all neighbours}\, n}
 \left(\begin{array}{ccc}
 k_x & 0 & 0\\
 0 & k_y & 0\\
 0 & 0 & k_z
 \end{array}\right)_{n}
 \left(\begin{array}{c}
 u_x \\ u_y \\ u_z
 \end{array}\right)_{n},
\label{eq-mukx}
\end{equation}
where $k_x$, $k_y$, $k_z$ and $u_x$, $u_y$, $u_z$ denote the spring constants and displacements in $x$, $y$ and $z$ direction for each nearest neighbour $n$, respectively.
Often the components of the spring constant are categorised into \emph{central} and \emph{non-central} terms, \emph{central} when acting along the dimension of their subscript, e.g., $k_x$ along the $x$-direction and \emph{non-central} otherwise. We can reduce the computational complexity of the problem by assuming that central and non-central force constants are identical. This turns all force constant matrices into scalars. After this reduction the three dimensions of the system are decoupled into three identical independent problems and solving any one solves the full system. This ``scalar" model, or ``isotropic Born model",\cite{AkiO98, BorH54} 
 can be written in its stationary form as
\begin{equation}
-\omega^2 m_{j} u_j = \sum_l k_{jl} (u_l - u_j)
,\label{eq-dyna}
\end{equation}
where $\omega$ is the frequency of vibration and $u_j(t)=u_j e^{i \omega t}$. In matrix notation, we have an eigensystem with eigenvalues $-\omega^2$,
\begin{equation}
 -\omega^2 \mathbf{U}=\mathbf{M}^{-1}\mathbf{K}\mathbf{U},
\label{eq-dynmat}
\end{equation}
where $\mathbf{M}^{-1}\mathbf{K}$ is called the dynamical matrix and, due to \emph{infinitesimal translational symmetry},\cite{Sri90} always obeys the sum rule $\sum_l (\mathbf{M}^{-1}\mathbf{K})_{jl} =0$. In the clean case, we have that all masses are equal to a constant $\overline{m}$ and all spring constants are $\overline{k}$. With these definitions, the frequencies range from $0$ to the largest possible frequency $\omega_\text{max}^2=12 \overline{k}/\overline{m}$ and $\omega^2$ will always be given in units of $[{\overline{k}/\overline{m}}]$.

\subsection{The disordered case}
\label{sec-disorder}

We are interested in introducing disorder into the system. From \eqref{eq-dynmat}, it is clear that this can be done (i) by allowing the masses to vary such that $m_j \in [ \overline{m}-\Delta m/2, \overline{m}+\Delta m/2]$ and (ii) by having random spring constants $k_{jl} \in [ \overline{k}-\Delta k/2, \overline{k}+\Delta k/2]$. For simplicity, we will use the \emph{uniform} mass and spring constant distributions with $\overline{m}=\overline{k}=1$ and restrict our investigation to the two cases of either pure mass or pure spring constant disorder. Note that this choice sets the units as well. The classical problem presented in Eq.\ \eqref{eq-mukx}, particularly its stationary form \eqref{eq-dyna}, is very similar to the tight-binding Schr\"{o}dinger equation for the three-dimensional Anderson model of localization\cite{And58} at energy $E$ such that 
$
 (E-\epsilon_j) \psi_j = - \sum_{l} t_{jl} \psi_{l}
\label{eq-anderson}
$, 
where the $l$ summation is over all nearest neighbours and $\epsilon_j$ and $t_{jl}$ denote the onsite and hopping energies, respectively.\cite{BraK03}
For the mass-disordered model with fluctuating masses $m_j$ one can obtain the transformation relations
\begin{equation}\label{transformation}
E \leftrightarrow 6-\omega^2, \qquad
\epsilon_j(E)\leftrightarrow \omega^2m_j=(6-E)m_j\, .
\end{equation}
As shown in Ref.\ \onlinecite{PinSR12}, we can then reuse many of the results for the Anderson model and infer the phase diagrams of localization-delocalization transitions for the vibrational mass-disorder model. In Fig.\ \ref{fig-phase}(a), we show the estimated mobility edges for the case of pure vibrational mass disorder based on transforming the related estimates of the mobility edges in the Anderson model.\cite{BulSK87,GruS95}
The phase diagrams for the vibrational case are intriguing in many respects.\cite{PinSR12} First of all (i) there is clear evidence for delocalization-localization transitions due to disorder. Next, (ii) the strong disorder limits of $|2\Delta m| > \overline{m}$, with the possibility of negative masses, or $|2\Delta k| > \overline{k}$, with similarly possible negative spring constants, give rise to locally unstable regions (although globally stable) corresponding to negative $\omega^2$ solutions. Such modes are known in liquids as unstable instantaneous normal modes and are related to the relaxation dynamics of the liquids.\cite{MadK93} (iii) The separation of extended and localized states continues into these regions and hence do the transitions and (iv) there is a re-entrant behaviour for $\omega>0$ and $\Delta m$ ($\Delta k$) $< 2$. These extraordinary mobility edges and hence the phase diagrams have been confirmed by direct high-precision numerics.\cite{PinR11,PinSR12}

\begin{figure*}[p]
\begin{center}
\begin{minipage}[t]{0.01\textwidth}
\vspace{190pt}
(a)
\end{minipage}
\hspace*{-20pt}
\begin{minipage}[t]{0.245\textwidth}
\vspace{60pt}
\includegraphics[width=0.98\textwidth]{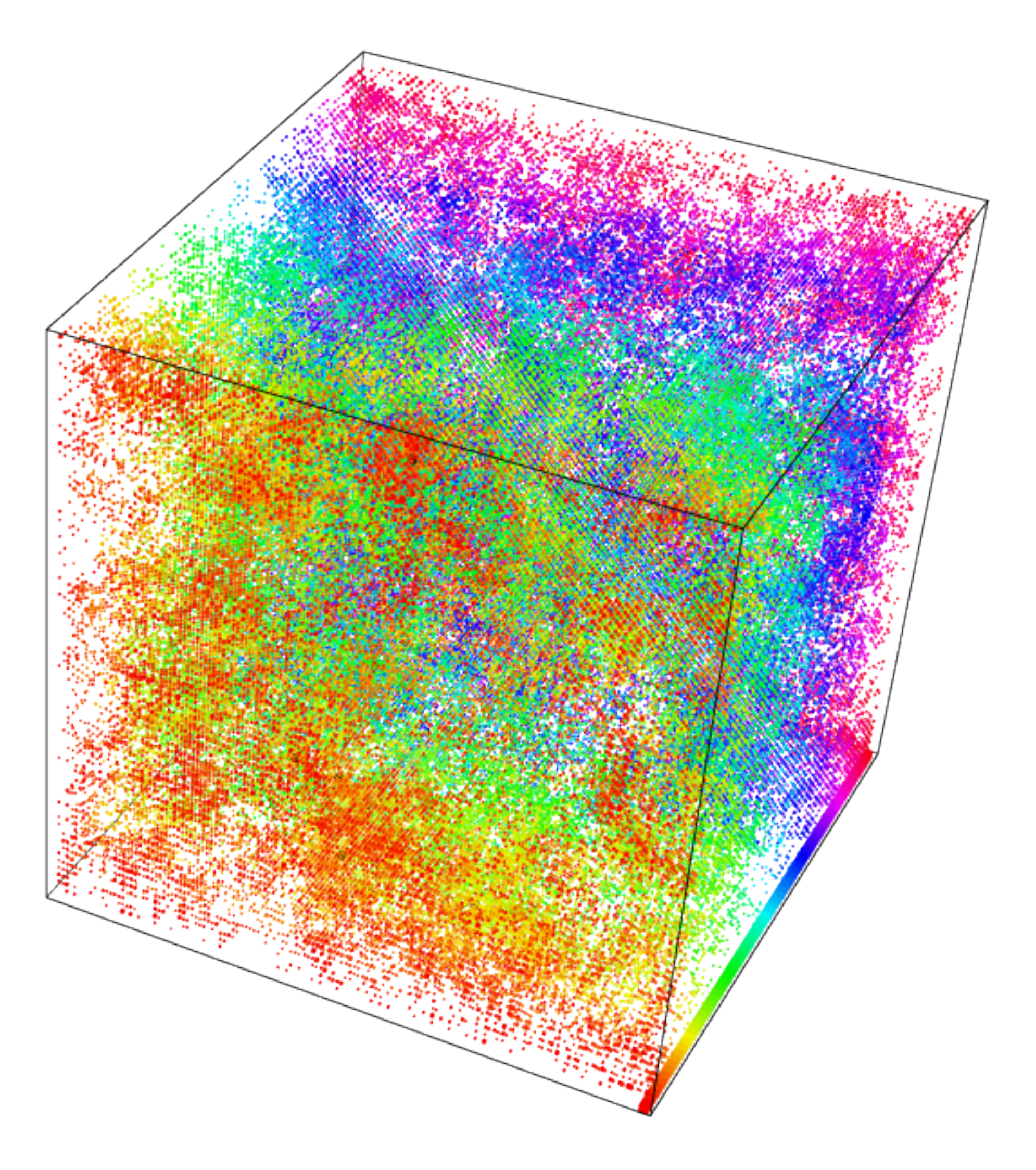}
\end{minipage}
\begin{minipage}[t]{0.01\textwidth}
\vspace{190pt}
(b)
\end{minipage}
\hspace*{3pt}
\begin{minipage}[t]{0.47\textwidth}
\vspace{0pt}
\includegraphics[width=0.98\textwidth]{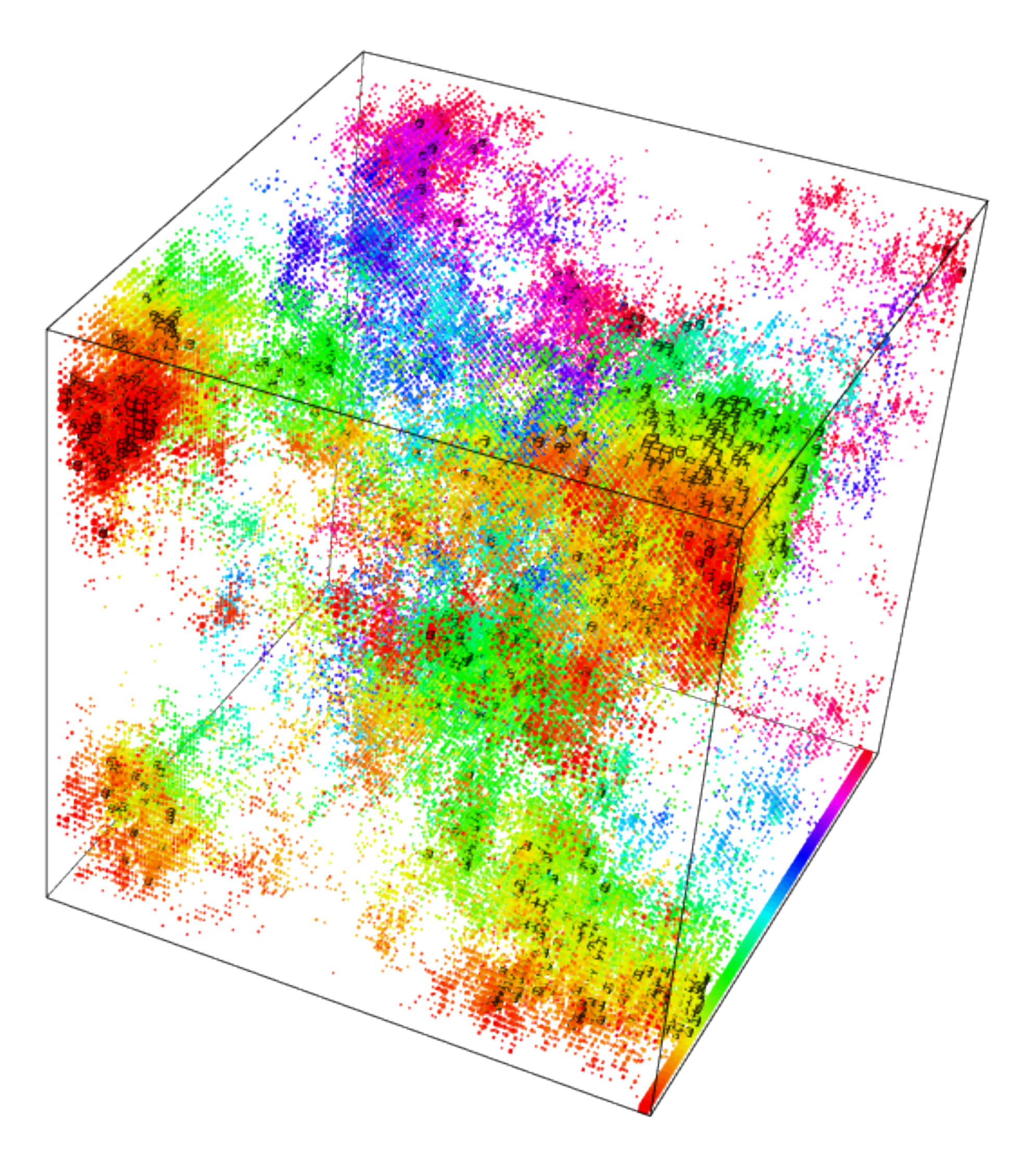}
\end{minipage}
\begin{minipage}[t]{0.01\textwidth}
\vspace{190pt}
(c)
\end{minipage}
\hspace*{-20pt}
\begin{minipage}[t]{0.245\textwidth}
\vspace{60pt}
\includegraphics[width=0.98\textwidth]{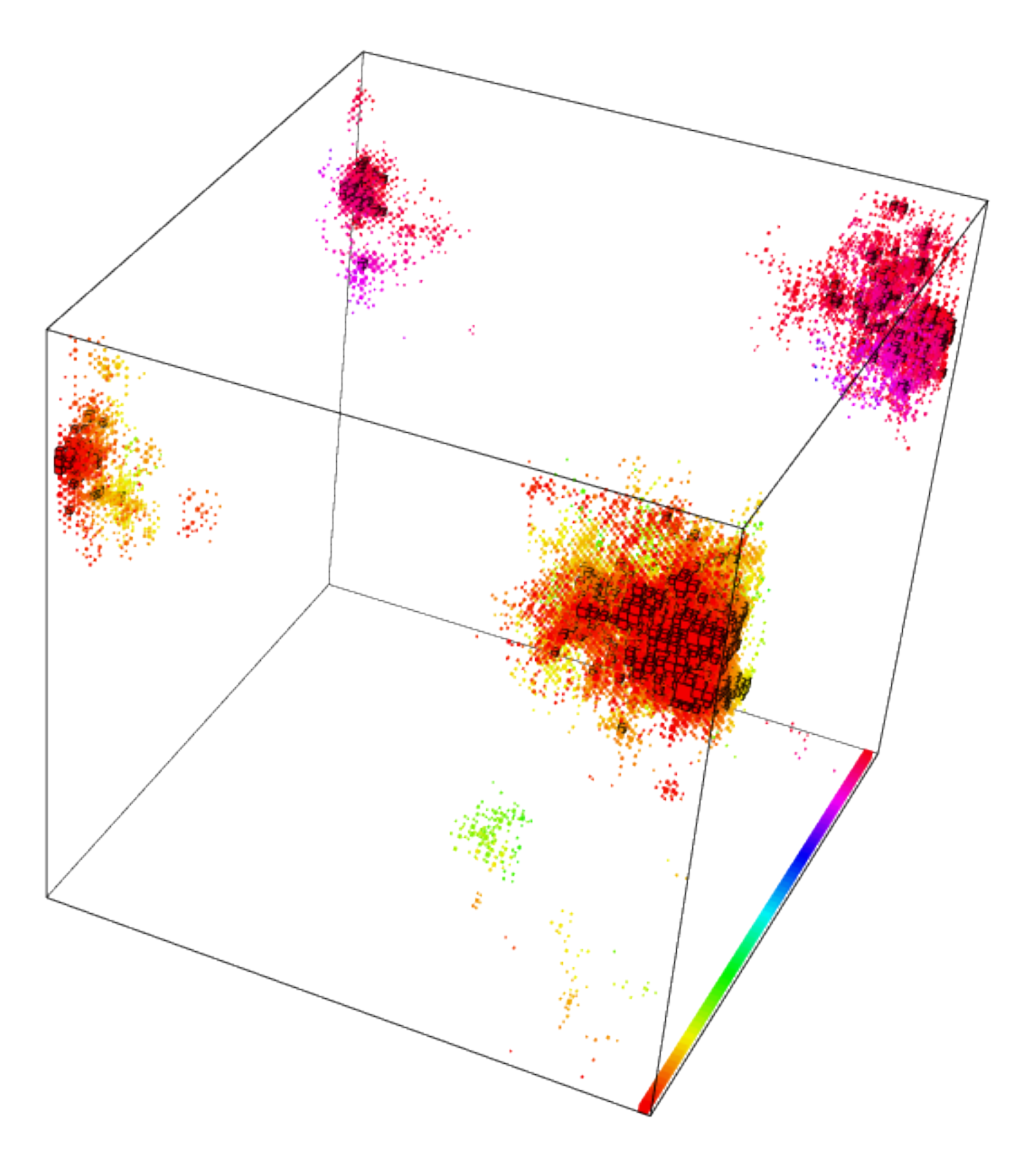}
\end{minipage}
\end{center}
\vspace*{-10pt}
\caption{(Color online) Schematic representation of amplitude distributions $|u_j|$ obtained from exact diagonalization for system of length $L^3=70^3$ for mass disorder $\Delta m=4$ and frequencies (a) $\omega^2 = 3$, (b) $\omega^2 = 4.5$ and (c) $\omega^2 = 6$. All sites with ${u(\vec{r}_j)}/L^{3}{\sum_{j} u(\vec{r}_j)} >1$ are shown as small cubes and those with black edges have ${u(\vec{r}_j)}/L^{3}{\sum_{j} u(\vec{r}_j)} > \sqrt{1000}$. The color scale distinguishes between different slices of the system along the axis into the page.\label{fig-Mstates}}
\end{figure*}
\begin{figure*}[p]
\begin{center}
\begin{minipage}[t]{0.01\textwidth}
\vspace{190pt}
(a)
\end{minipage}
\hspace*{-20pt}
\begin{minipage}[t]{0.245\textwidth}
\vspace{60pt}
\includegraphics[width=0.98\textwidth]{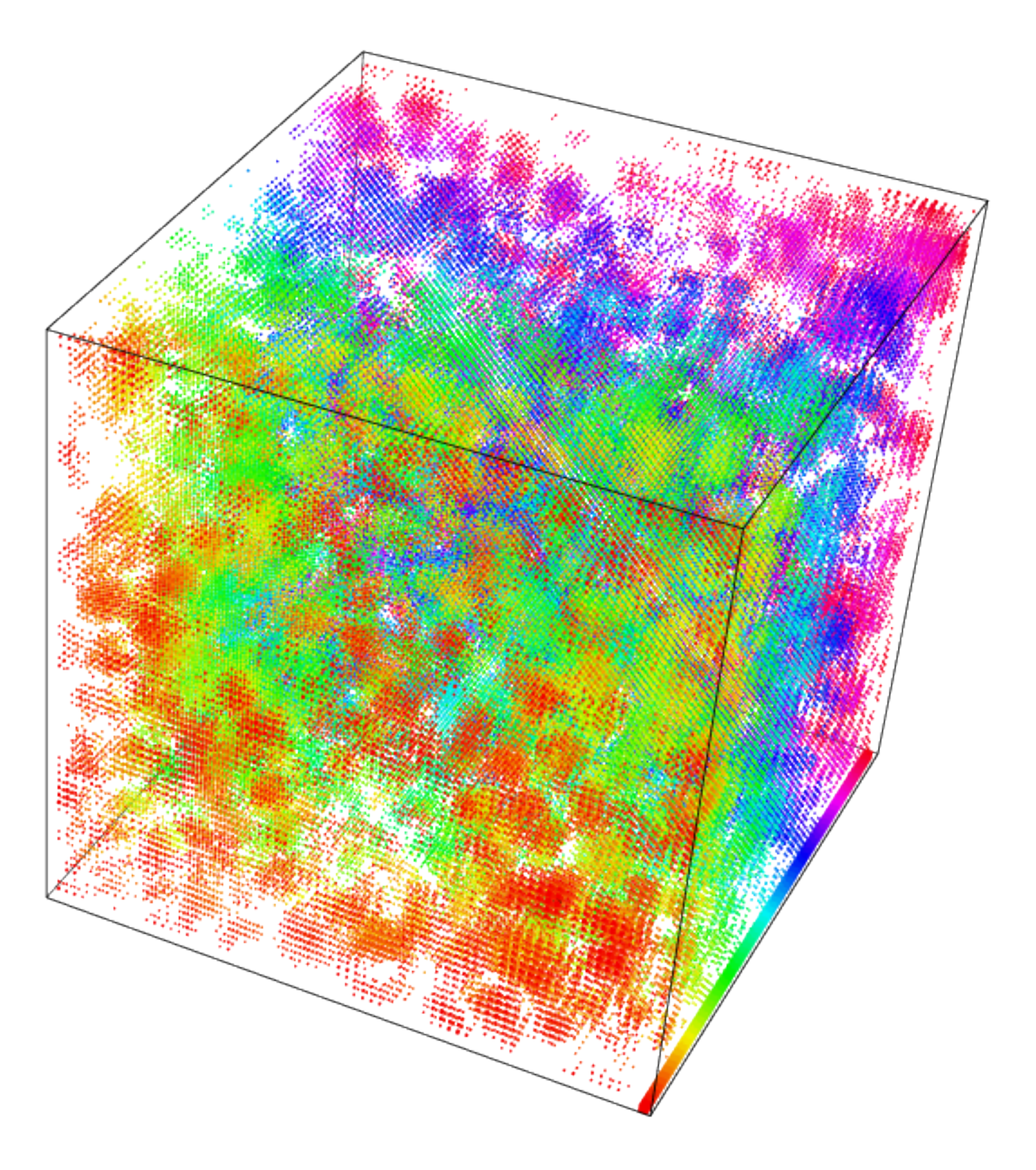}
\end{minipage}
\begin{minipage}[t]{0.01\textwidth}
\vspace{190pt}
(b)
\end{minipage}
\hspace*{3pt}
\begin{minipage}[t]{0.47\textwidth}
\vspace{0pt}
\includegraphics[width=0.98\textwidth]{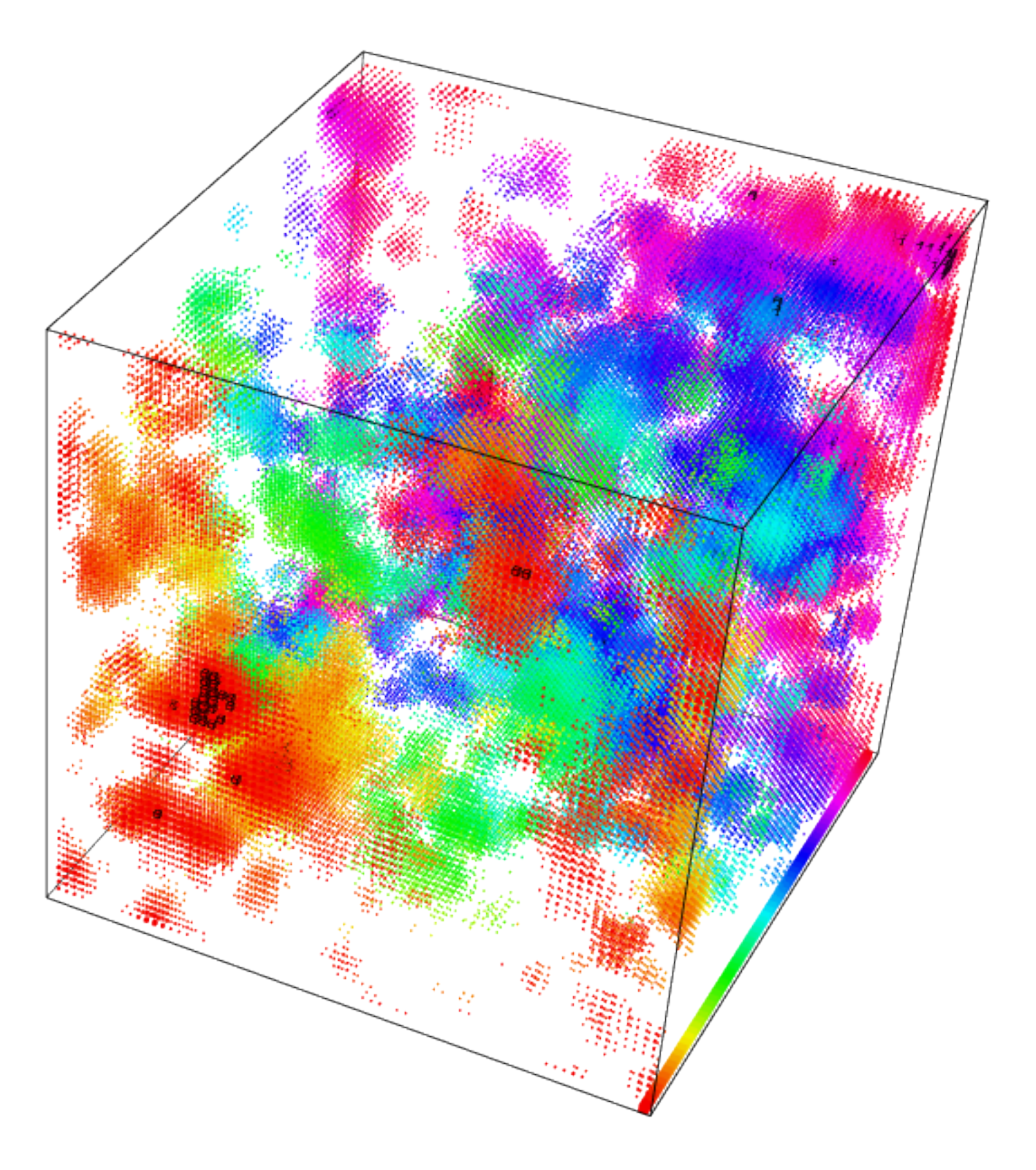}
\end{minipage}
\begin{minipage}[t]{0.01\textwidth}
\vspace{190pt}
(c)
\end{minipage}
\hspace*{-20pt}
\begin{minipage}[t]{0.245\textwidth}
\vspace{60pt}
\includegraphics[width=0.98\textwidth]{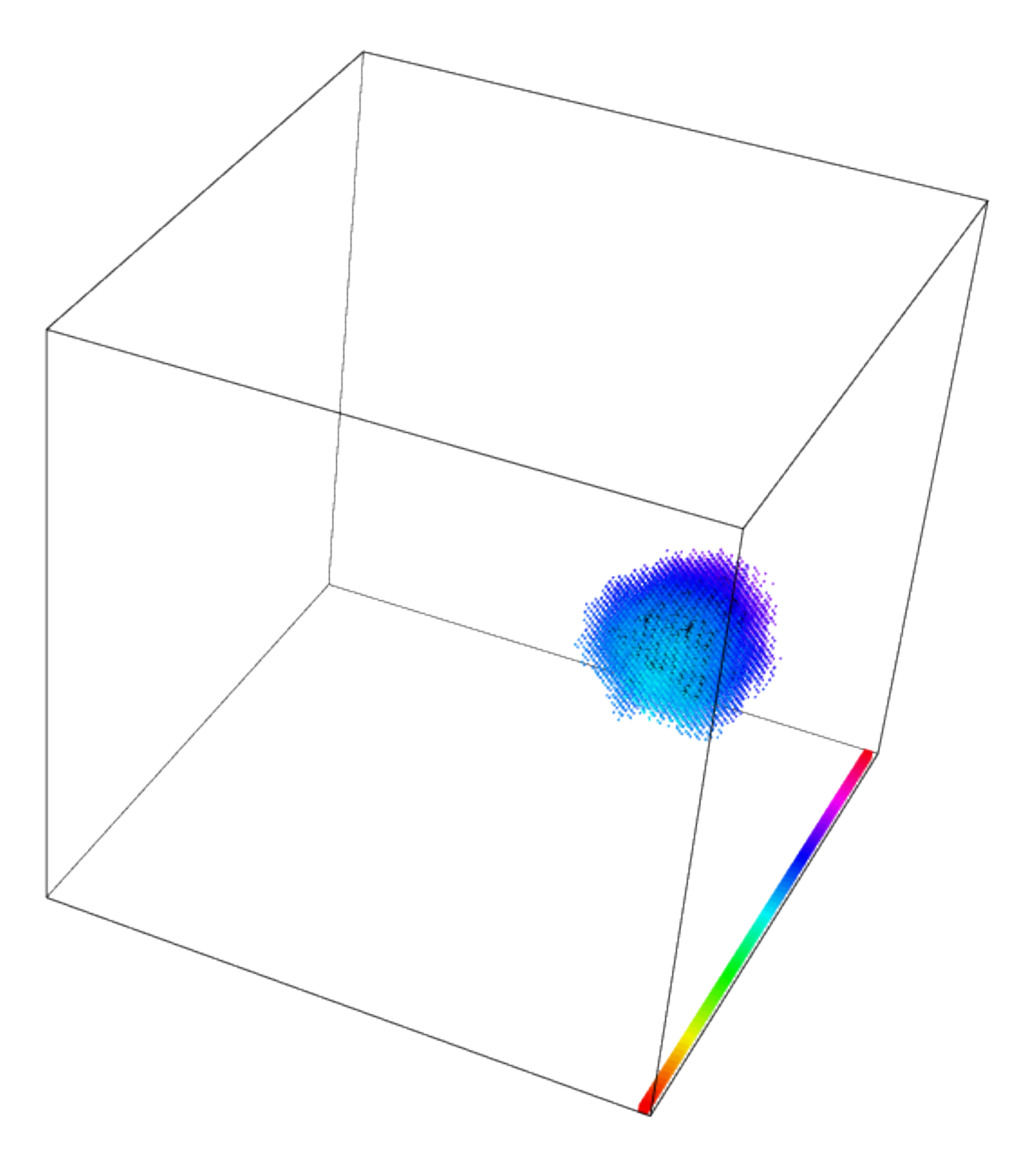}
\end{minipage}
\end{center}
\vspace*{-10pt}
\caption{(Color online) Schematic representation of amplitude distributions $|u_j|$ as in Fig.\ \ref{fig-Mstates} obtained for spring disorder $\Delta k=1$ and frequencies (a) $\omega^2 = 12$, (b) $\omega^2 = 12.5$ and (c) $\omega^2= 13.03$. The size of the cubes and their colour is chosen as in Fig.\ \ref{fig-Mstates}.\label{fig-Kstates}}
\end{figure*}

\section{Localization properties of eigenstates}
\label{sec-loceig}

\subsection{Numerical diagonalization}
\label{sec-ND}

Let us start our investigation of \eqref{eq-dynmat} by looking at some typical eigenstates obtained by exact diagonalization. In particular, we are using a combination of the iterative numerical eigensystem packages {\sc Arpack}\cite{LehSY98} and {\sc Pardiso}.\cite{SchBR06} We find this combination to be most effective when dealing with both the unsymmetric and the symmetric cases of pure mass and spring disorder, respectively.\cite{Note1}

In Fig.\ \ref{fig-Mstates}, we show eigenstates for the pure mass disorder case corresponding to three eigenfrequencies which lie in regions which according to the phase diagram (Fig.\ \ref{fig-phase}(a)) should be extended, close to the mobility edge and localized. We see from Fig.\ \ref{fig-Mstates} that these characterisations reflect the apparent nature of these vibrational states. For Fig.\ \ref{fig-Mstates}(a), the local amplitude of vibrations at each site is roughly of similar magnitude throughout the system, whereas for Fig.\ \ref{fig-Mstates}(c), the vibrations are confined to a small region in the cube. Figure \ref{fig-Mstates}(b) displays the characteristic properties of a \emph{critical} wave function at the Anderson mobility edge.\cite{VasRR08}

For the pure spring disorder case as in Fig.\ \ref{fig-Kstates}, we see that the vibrations for the three shown frequency values may also be classified into extended, critical and localized classes. This classification indeed agrees with the computed phase diagram as shown in Fig.\ \ref{fig-phase}(b) for the pure spring disorder case. However, we also see that the character of the states seem subtly different from the pure mass disorder ones. The vibrations seem to be more around certain vibration centres and radiate outward roughly symmetrically from these centres.\cite{LudTED05} Although not the topic of the present investigation, we emphasise that this should make the multifractal analysis of such states very informative, in particular its comparison with the recently proposed symmetry of the multifractal spectrum.\cite{LudSTE01,VasRR08,MirFME06}

\subsection{Vibrational density of states}
\label{sec-vdos}

In order to numerically obtain the VDOS, the computation of \emph{all} states is required. The iterative methods applied in section \ref{sec-ND} are then no longer efficient and we employ a standard {\sc LaPack}\cite{AndBBB87} dense matrix routine ({\sc DGEEV}). 

\begin{figure}[tb]
(a)\includegraphics[width=0.45\textwidth]{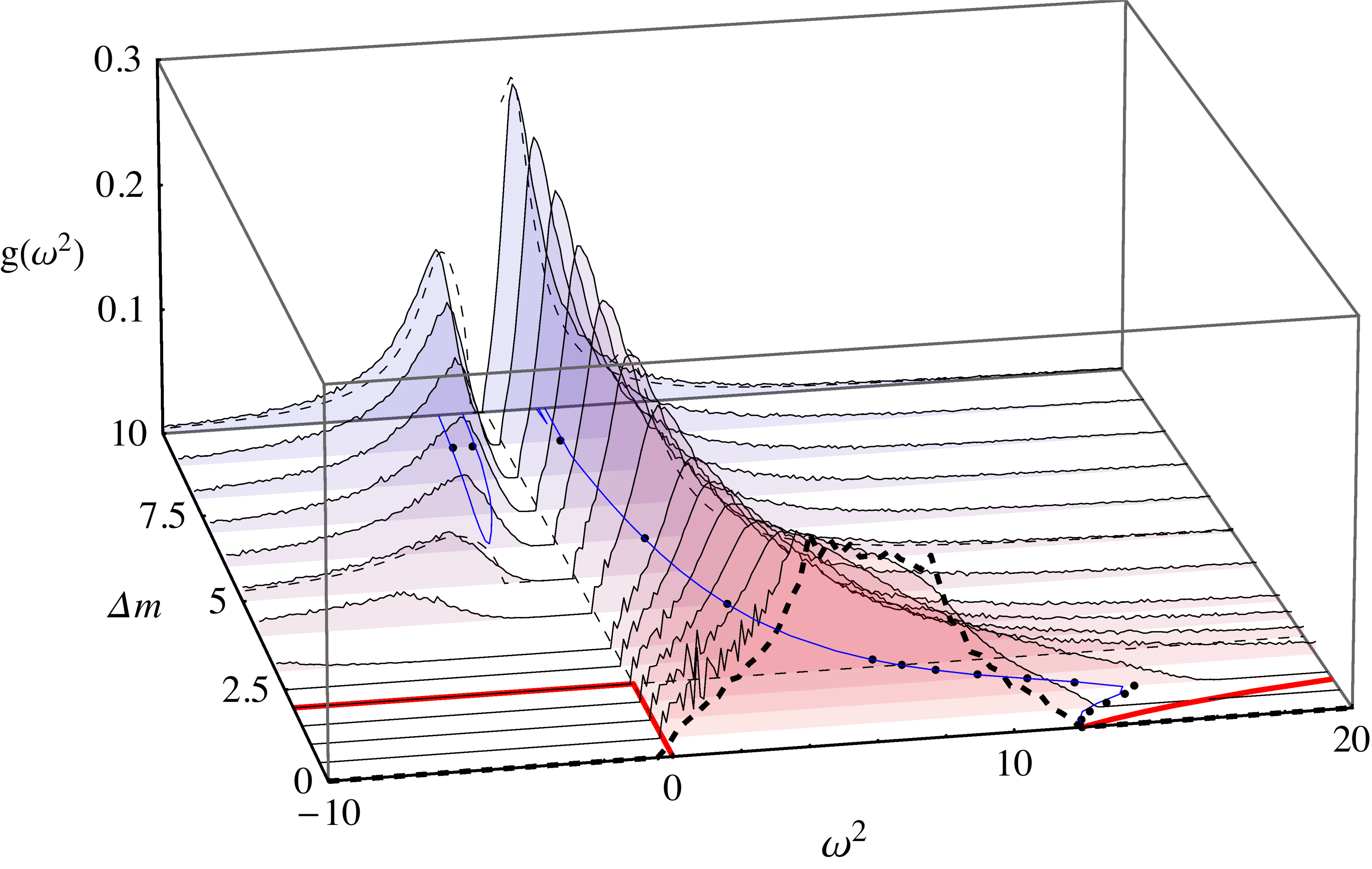} 
(b)\includegraphics[width=0.45\textwidth]{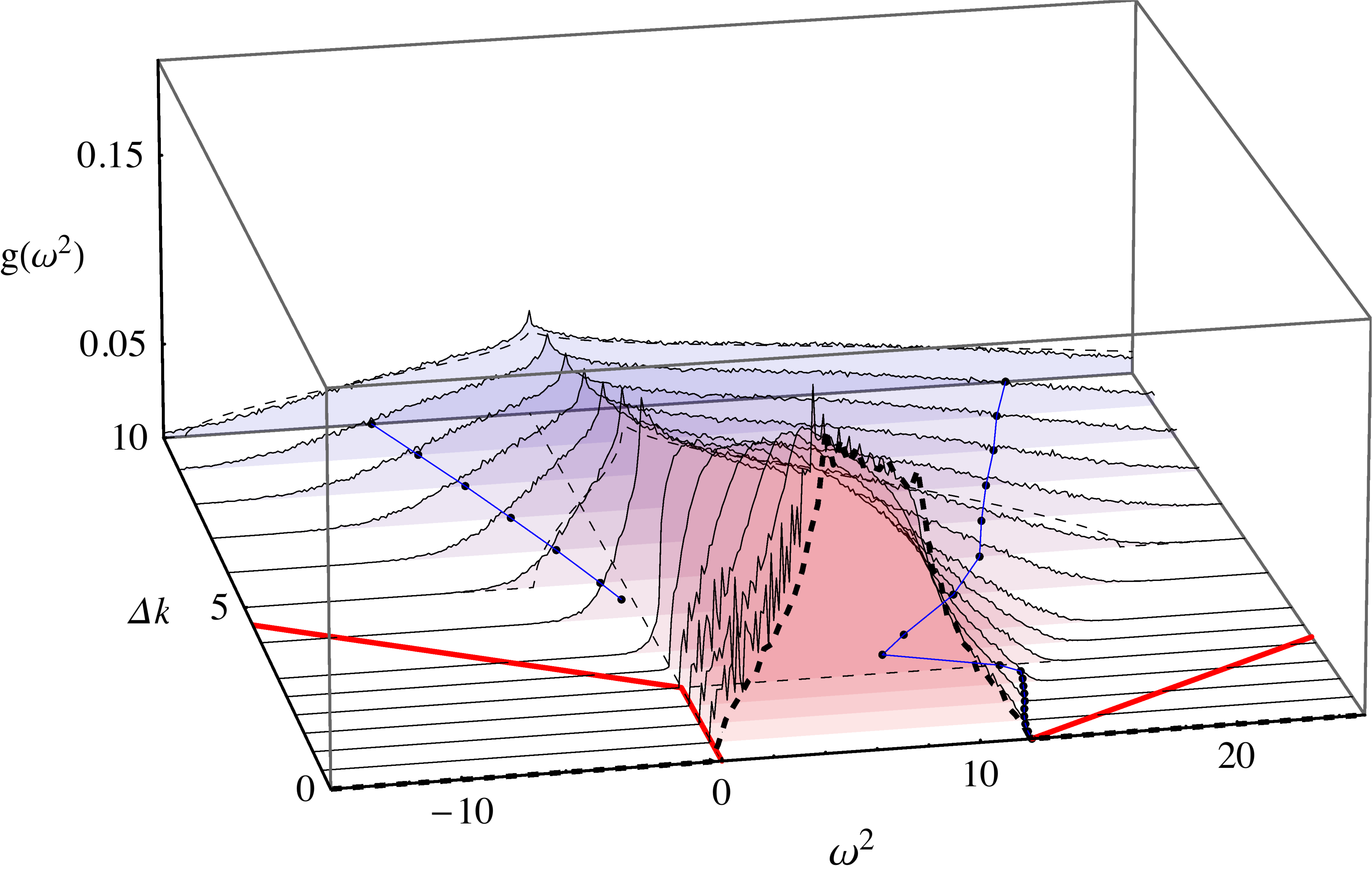} 
\caption{(Color online) Vibrational density of states $g(\omega^2)$ as a function of frequency $\omega^2$ and various disorders of (a) $\Delta m$ and (b) $\Delta k$. The blue and red lines in the base denote the trajectories of the localization-delocalization transition and the band edges, respectively, see Fig.\ \ref{fig-phase}. The thin dashed lines for $\Delta m, \Delta k= 5, 10$ are results from CPA calculations (see Appendix \ref{sec-cpa} for details). The thick dashed line is the clean simple cubic density of states and identical in both cases.\label{fig-bosepeaks}\label{fig-vdos}}
\end{figure}

We have calculated the VDOS $g(\omega^2)=g(\omega)/2\omega$ for disorders $\Delta m, \Delta k = 0.5, 1, 1.5, 2, 2.5, 3, 4, 5, 6, 7, 8, 9$ and $10$ for cubes with volume $L^3= 15^3$ for $50$ disorder configurations. This results in roughly $170,000$ $u_j$'s for each disorder. In Figs.\ \ref{fig-bosepeaks}(a) and \ref{fig-bosepeaks}(b) we show the results for $g(\omega^2)$ as a functions of $\omega^2$ for all mass and spring constant disorder magnitudes respectively. 
We find for both types of disorder that the van Hove singularities in the VDOS become smeared out upon increasing the disorder. In addition, there are the usual low-frequency peaks corresponding to standing waves in the simulation box. These peaks indicate the presence of plane-wave-like states.\cite{LeoTWB06, LeoBTW05, MonM09} We perform analytical calculations of the VDOS using the coherent-potential approximation (CPA, see Appendix A).\cite{YonM73} Except for the standing-wave peaks (which are absent in the $L\rightarrow\infty$ CPA calculations) there is very good agreement between the analytical and numerical results as can be seen from Figs.\ \ref{fig-vdos}(a) and \ref{fig-vdos}(b). Using the CPA one can easily evaluate the maxima of the ``reduced VDOS" $g(\omega)/\omega^2=2 g(\omega^2)/\omega $ (``boson peaks''). For small disorder ($\Delta m < 1$, $\Delta k < 1$) these peaks are identical with the transverse van Hove singularities, located at $\omega^2=4$. For larger disorder the BPs become disorder-dominated and no longer reflect the underlying lattice symmetry. This can be (and has been) checked by CPA calculations using a Debye Green's function $G_0(z)=\int_{-\infty}^\infty g_D(\lambda)/(z-\lambda)$ with $\lambda=\omega^2$, $g_D(\omega^2)=3\omega/2\omega_D^3$. In these calculations the BP positions for $\Delta m>1$, $\Delta k>1$ coincide with those of the lattice calculations. It has been shown in Ref. \onlinecite{SchDG98} that the BP separates a nearly plain wave regime from a regime where disorder is dominant (random-matrix regime). We find from analysing our VDOS data that this is also the case for our model systems. 

However the scenario for mass and spring-constant disorder is very different. In the spring-constant disorder case the BP, and with it, the range of nearly plane waves goes continuously towards zero near $\Delta k = 2.5$. In CPA there are no states with $\omega^2<0$ below this value. In the mass disorder case the BPs and correspondingly the low-frequency range of nearly plane waves extend towards $\Delta m \rightarrow \infty$. This can be easily understood by the transformation rule \eqref{transformation}, which states that the mass fluctuations are suppressed by a factor $\omega^2$. Therefore for $\omega^2\rightarrow 0$ there are always plane waves in the infinite-volume system, which are converted to standing waves at finite volume. 

It is interesting to note that in the mass disorder case a peak on the negative $\omega^2$ side develops for high values of $\Delta m$ near the $\omega^2<0$ mobility edge. On the positive $\omega^2$ side both the peak in $g(\omega^2)$, the BP and the mobility edge approach each other with increasing $\Delta m$. This confirms that there is no proportionality between $\omega_{\rm BP}$ and $\omega_{\rm c}$ as postulated in Ref.\ \onlinecite{KanRB01}. The absence of such a simple relationship was also already discussed in
Refs.\ \onlinecite{ScoSAA06} and \onlinecite{TarLNE02}.
\subsection{Participation numbers}
\label{sec-PR}

The participation number $P_L(\omega_n)$ is a measure of the number of sites in the lattice that are contributing to the vibrational excitation of the $n$th vibrational eigenstate $u_1(n),u_2(n),\ldots,u_{L^3}(n)$. It can be defined as\cite{CanV85}
\begin{equation}
 P_L^{-1}(\omega_n) = {L^3 \sum_{j} u_{j}^4(n)}
\end{equation}
in analogy with the electronic case. We emphasise that the normalisation $\sum_{j}u_{j}^2(n)=1$, automatically observed for electronic eigenstates by the Born rule, has to be enforced for the vibrational case for consistency in the comparison between different eigenstates.\cite{LudSTE01} A fully extended vibration will lead to $P_L(\omega_n)=1$ whereas a vibration localized at a single site corresponds to $P_L(\omega_n)=1/L^3$ and hence $0$ in the limit $L \rightarrow \infty$.

\begin{figure}[tb]
(a)\includegraphics[width=0.45\textwidth]{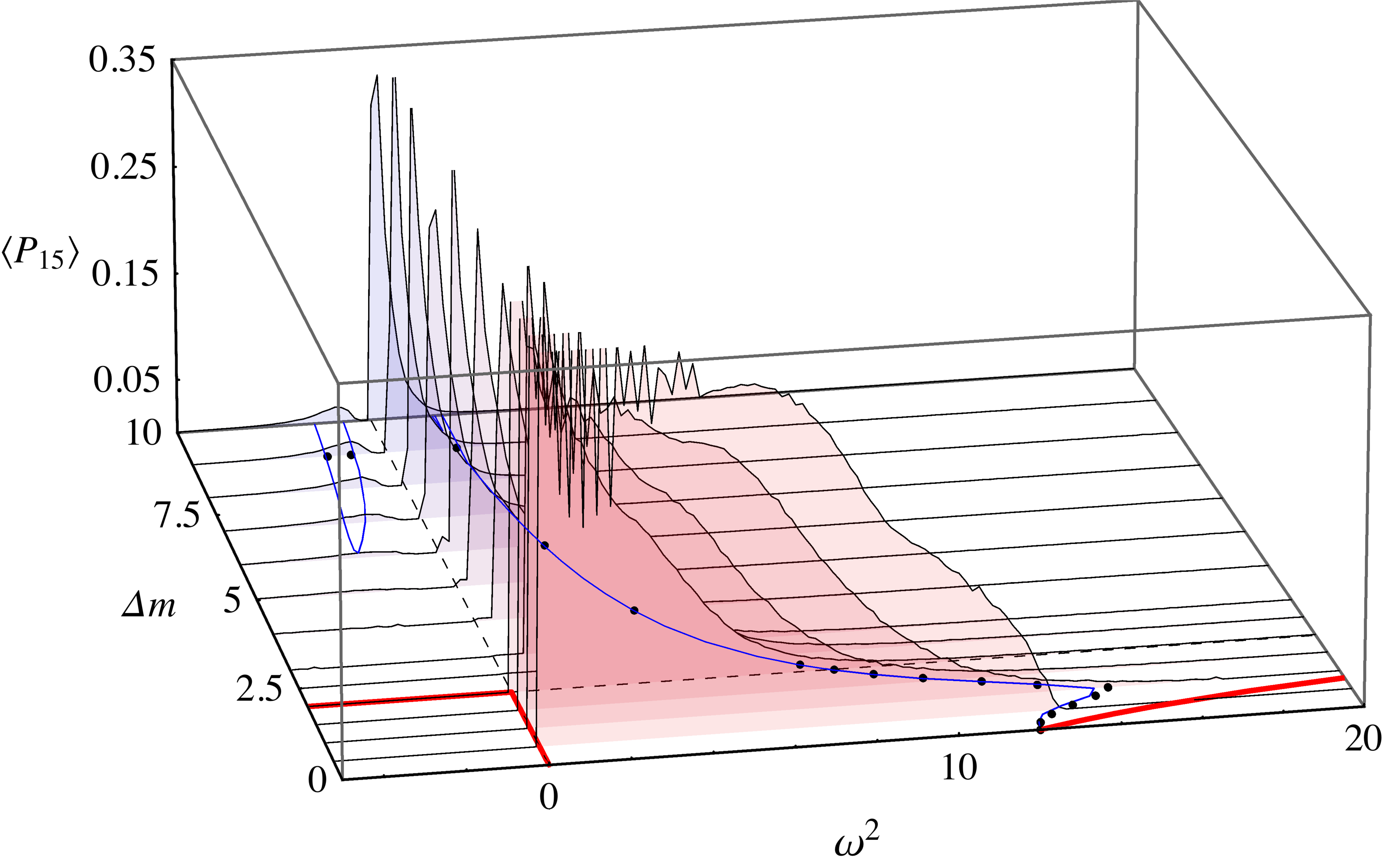} 
(b)\includegraphics[width=0.45\textwidth]{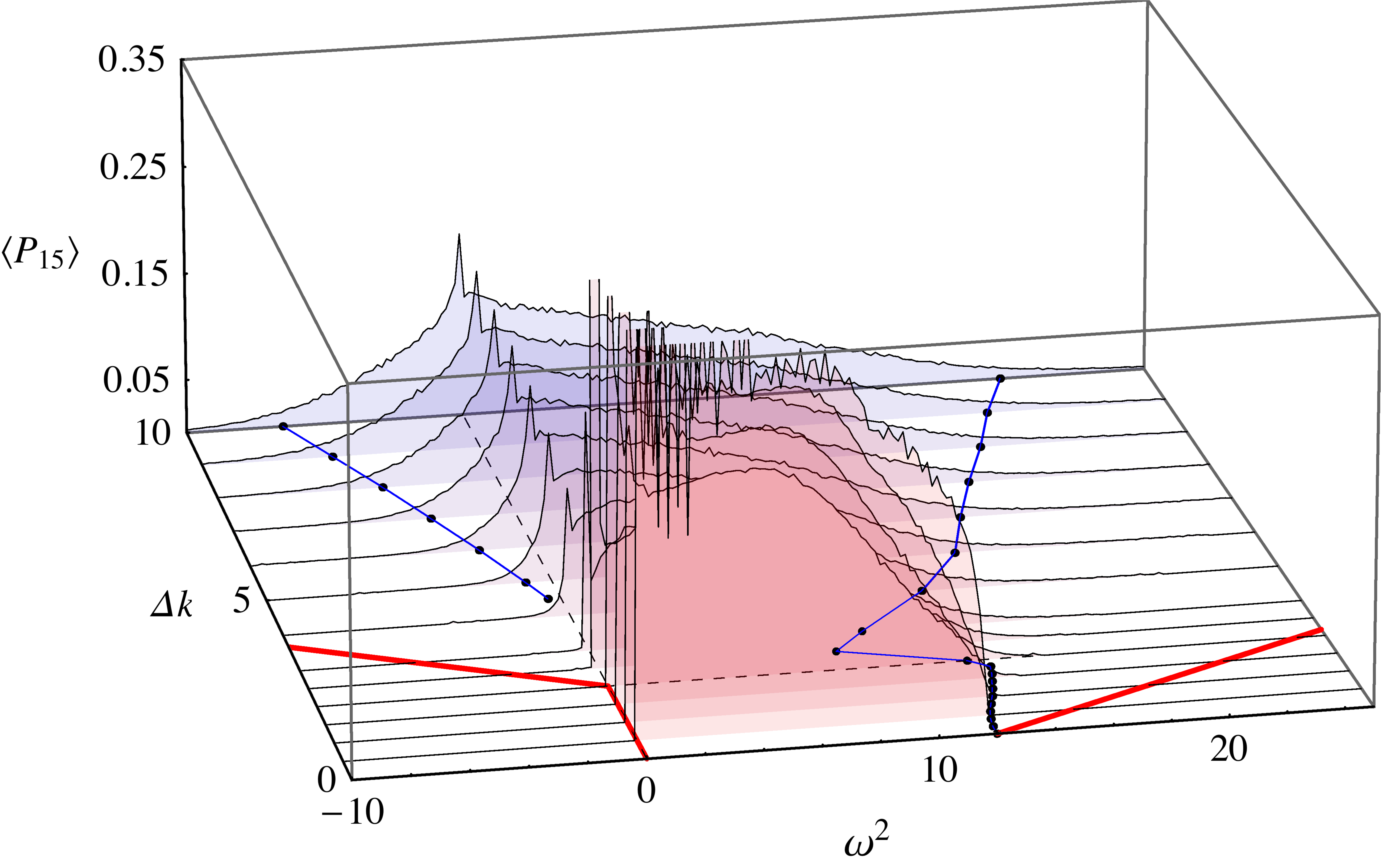} 
\caption{(Color online) Participation numbers $P_{15}$ as a function of $\omega^2$ for various (a) mass disorders $\Delta m$ and (b) spring constant disorders $\Delta k$, averaged over $50$ disorder realisations. The dotted grey and red lines in the base denote the phase boundaries and the band edges respectively as in Fig.\ \ref{fig-phase}.\label{fig-partratios}}
\end{figure}

We average the participation numbers in discrete frequency intervals over $50$ disorder realizations and plot them for each disorder at $L=15$ in Figs.\ \ref{fig-partratios}(a) and \ref{fig-partratios}(b) for mass and spring disorder, respectively. We find that the transition from delocalized to localized behaviour as found in section \ref{sec-TMM_res} does not lead to a clear crossing of $P_L$ for system sizes $L=5$, $10$ and $15$. We expect to see such a crossing only when going to much larger system sizes and upon increasing the number of disorder samples. Thus our results show the difficulties associated with the use of participation numbers in studying the present transition in agreement with a recent attempt by Monthus et al.\cite{MonG10}

In general, the results for $P_L$ nevertheless confirm the phase diagrams presented in Figs.\ \ref{fig-phase}(a) and \ref{fig-phase}(b) as the extended regions of the phase diagrams are matched with states of higher participation. The VDOS results of Fig.\ \ref{fig-vdos} are also confirmed qualitatively as extended states usually lead to higher $P_L$ values than localized ones. In particular, we note the emergence of finite $P_L$ values in the negative $\omega^2$ regime for large mass disorder as well as the pronounced tail in the same frequency regime for strong spring constant disorder.

\subsection{Vibrational Eigenstate Statistics}
\label{sec-PTS}

\begin{figure}[tb]
(a)\includegraphics[width=0.45\textwidth]{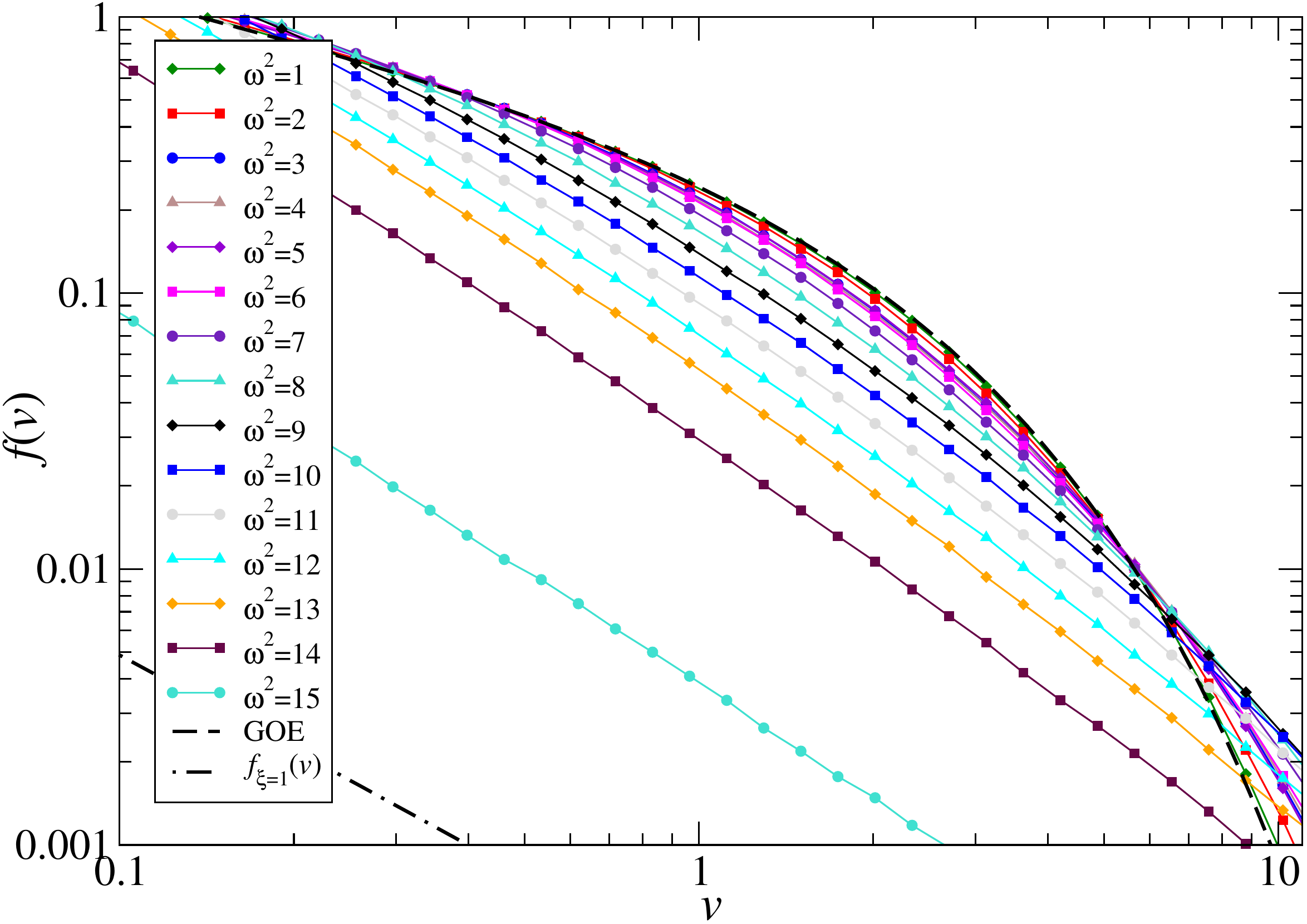} 
(b)\includegraphics[width=0.45\textwidth]{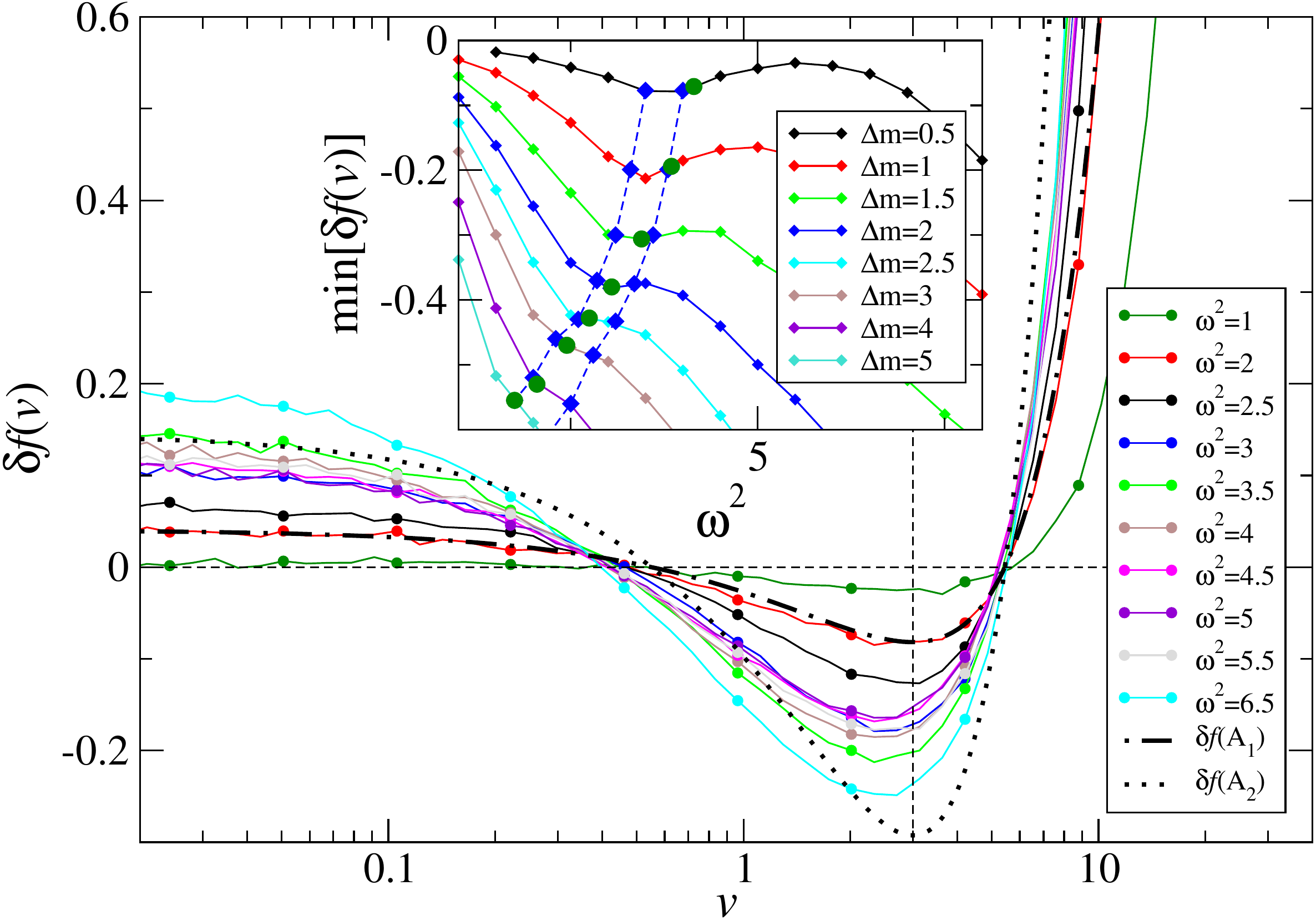} 
\caption{(Color online) For disorder $\Delta m=1$ (a) $f(v)$ for a range of frequencies as labelled in the figure, with the pure GOE result indicated by the dashed line, and the highly localized result indicated by the dot-dashed line. (b) $\delta f(v)$ for a range of frequencies where every 5$^{\text{th}}$ data point has a symbol. GOE here corresponds to the $\delta f=0$ line whereas the vertical $v=3$ line is the minimum position of the analytical $\delta f$.  The dot-dashed and dotted lines indicate $\delta f$ of \eqref{eq-perturbGOE} with constants $A_1=0.0545$ and $A_2=0.195$, respectively. Inset: the value of the minimum of $\delta f$ plotted as a function of $\omega^2$ for different mass disorders. The green dots correspond to the positions of the BP obtained via CPA calculations and the dashed blue lines estimate the position of the minima and their width is equal to the separation of the minimum function data points, all shown in the phase diagram in Fig.\ \ref{fig-phase}(a).\label{fig-PTS}}
\end{figure}

\begin{figure}[tb]
(a)\includegraphics[width=0.45\textwidth]{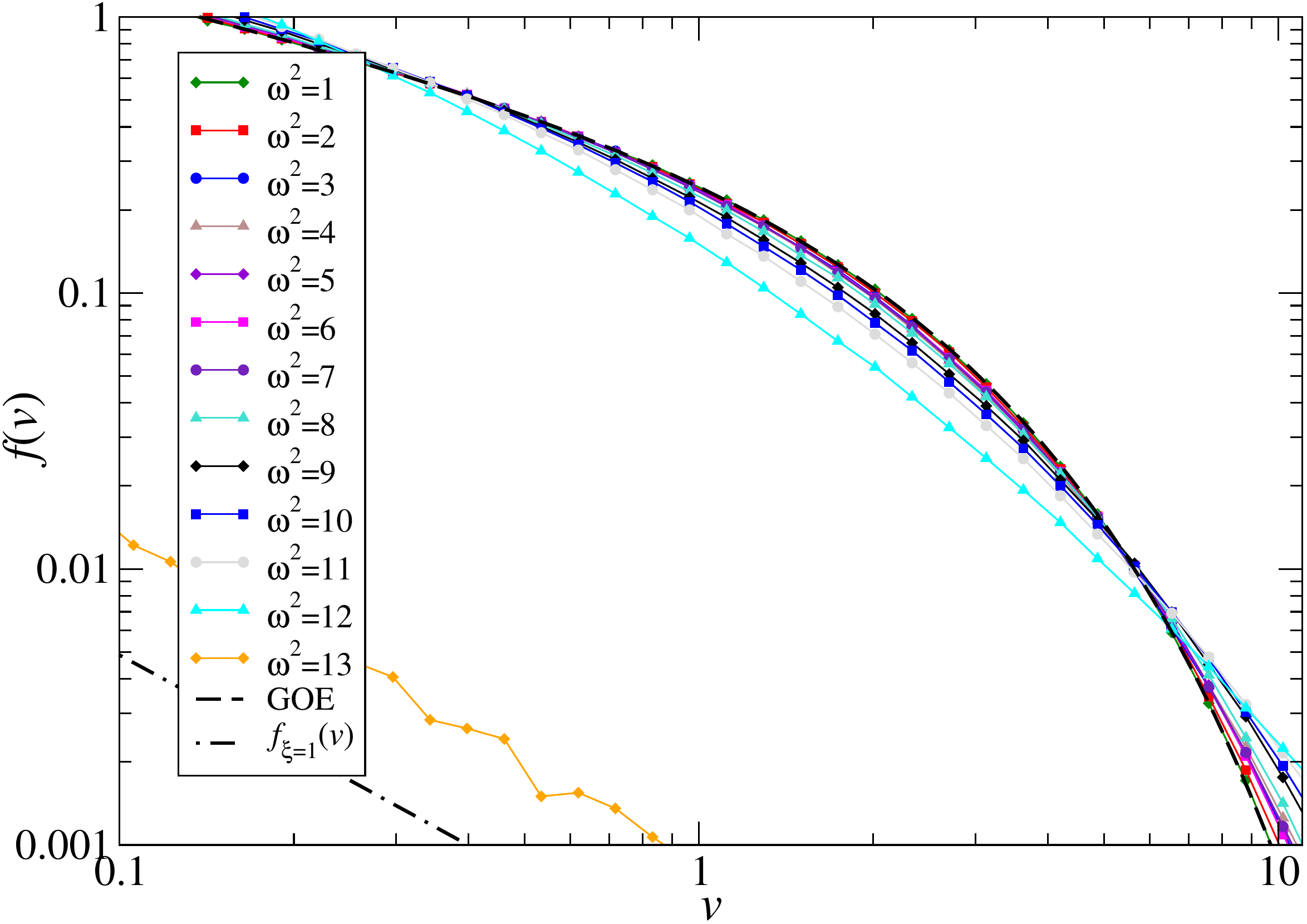} 
(b)\includegraphics[width=0.45\textwidth]{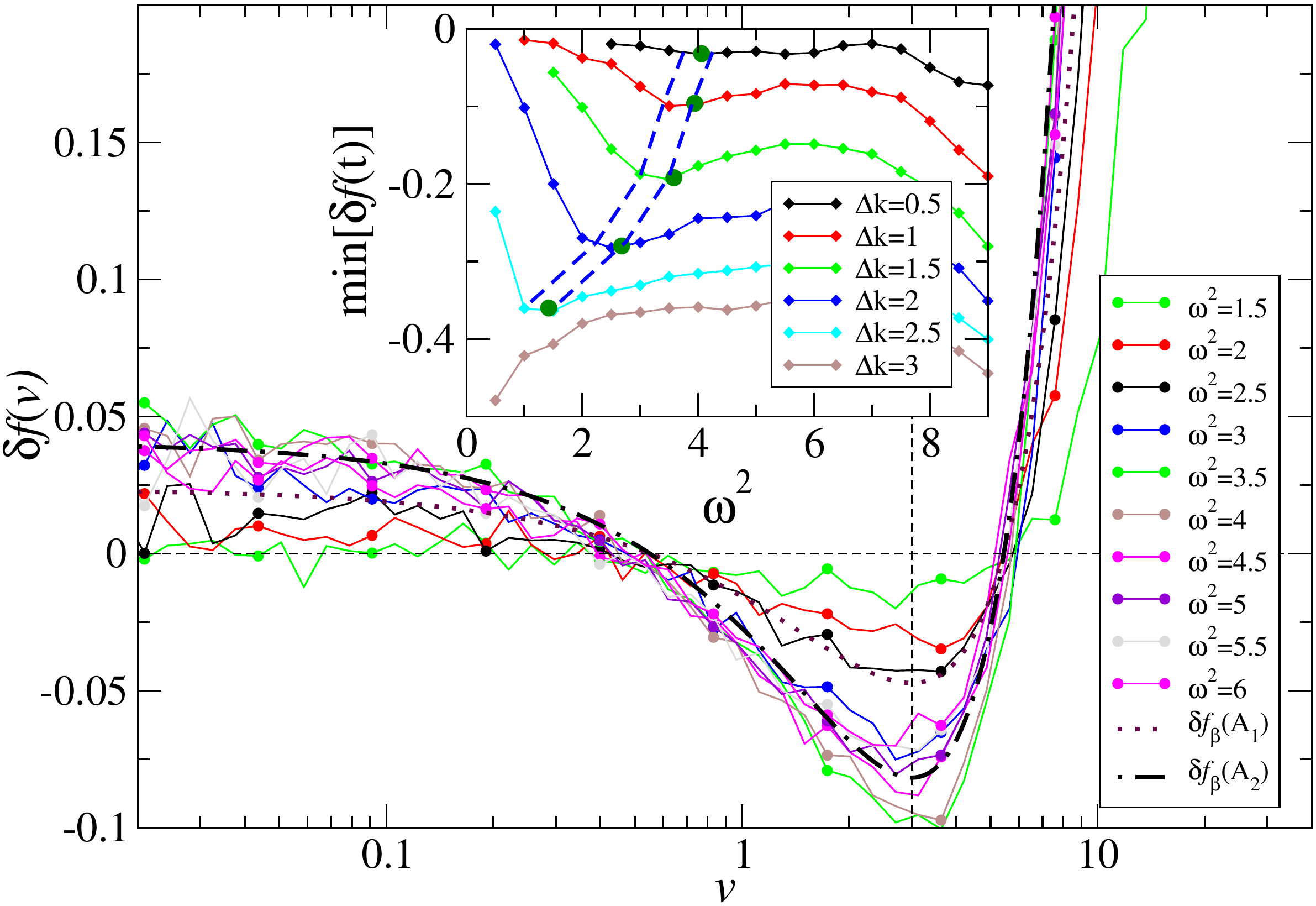} 
\caption{(Color online) For disorder $\Delta k=1$ (a) $f(v)$ for a range of frequencies as labelled in the figure with the pure GOE and fully localized behaviours as indicated in Fig.\ \ref{fig-PTS}(a). Panel (b) shows $\delta f(v)$ for a range of frequencies where every 5$^{\text{th}}$ data point has a symbol. GOE here corresponds to the $\delta f=0$ line whereas the vertical $v=3$ line is the minimum position of the analytical $\delta f$.  The dot-dashed and dotted lines indicate $\delta f$ of \eqref{eq-perturbGOE} with constants $A_1=0.0315$ and $A_2=0.0545$, respectively. Inset: as in Fig.\ \ref{fig-PTS} also plotted in the corresponding phase diagram Fig.\ \ref{fig-phase}(b).\label{fig-PTS-k}}
\end{figure}

Disordered quantum systems exhibit irregular fluctuations of eigenfunctions, which can be studied from the statistics of the local amplitudes.\cite{Meh91,Haa92} In the universal regime (of mostly weak disorder), random matrix theory can classify these fluctuations into universality classes such as the Porter-Thomas distribution\cite{Por65} of the GOE.\cite{Dys62} Upon increasing the disorder, corrections to GOE have been studied which we expect to see present also in the case of our vibrational disorder.\cite{FyoM94,UskMRS00}
We determine the distribution function
\begin{equation}
f(v) = \frac{\Delta}{L^3} \left\langle \sum_{n,j} \delta\left(v-|u_j(n)|^2 L^3\right) \delta \left(\omega^2-\omega^2_n\right)\right\rangle
\end{equation}
where $\Delta$ is the mean level-spacing, $\langle \text{ } \rangle$ denotes an average over disorder realisations and the vibrational eigenvectors are normalised so that $\langle | u_j(n)|^2\rangle = L^{-3}$. 
%
For all disorders mentioned in section \ref{sec-vdos} we calculate $f(v)$ from over two million amplitudes at $L^3=70^3$ for frequencies throughout the phase diagram at intervals of $\delta\omega^2=0.5$ and plot them for mass and spring disorder in Figs.\ \ref{fig-PTS}(a)--\ref{fig-PTS-k}(a), respectively. We include the Wigner estimate from random matrix theory, $f^{W}_\text{GOE}=\text{exp}(-v/2)/\sqrt{2\pi v}$.\cite{Por65} For exponentially localized states, one finds $f_{\xi}(v)\sim\ln^2(c^2L^3/v)/v$ with $c$ a disorder dependent constant and $\xi$ is the localisation length.\cite{Nik01b,MulMMS97} We also include the theoretical result for a maximally localized scenario, where $\xi=1$ in Figs.\ \ref{fig-PTS}(a)--\ref{fig-PTS-k}(a). We see in Fig.\ \ref{fig-PTS}(a) that the curves for increasing mass disorder increasingly depart from GOE, whereas for spring constant disorder in Fig.\ \ref{fig-PTS-k}(a) there is an abrupt departure from GOE when the localization-delocalization transition is crossed.
In Figs.\ \ref{fig-PTS}(b) and \ref{fig-PTS-k}(b) we plot the relative difference $\delta f$ between $f$ and $f^{W}_\text{GOE}$ as
\begin{equation}
\delta f(v) = \frac{f(v)}{ f^{W}_\text{GOE}}-1,\label{eq-diffGOE}
\end{equation}
and include the analytical estimate of departure from GOE as derived for the electronic Anderson model \cite{FyoM94}
\begin{equation}
\delta f (v) \backsimeq A \left( \frac{3}{4} - \frac{3v}{2} + \frac{v^2}{4} \right),
\label{eq-perturbGOE}
\end{equation}
where $A$ is a constant related to the diffusion in the system.  We see that for small frequencies the analytical estimate is very well suited to our data and we show that for the mass disorder case a value of $A_1=0.0545$ has a good fit for $\delta f$ of $\omega^2=2$ and similarly $A_1=0.0315$ has a good fit for $\delta f$ of $\omega^2=2.5$ in the spring constant disorder case. For higher frequencies this fit continues in the spring constant disorder case, where for a value $A_2=0.0545$ we have a good agreement with $\delta f$ of $\omega^2=6$. This is not the case in the mass disorder case where the minimum values of $\delta f$ shift from $v=3$ and as an illustration we show that for $A_2=0.195$ the difference $\delta f$ fits the $\omega^2=3.5$ results only for small $v$ but very quickly deviates for increasing $v$.

Upon further increasing $\omega^2$, we see that there is again a region where the agreement with $f^W_\text{GOE}$ becomes better. This behaviour has not previously been observed (neither in the electronic case nor in calculations on vibrational modes).

In the inserts of Figs.\ \ref{fig-PTS}(b) and \ref{fig-PTS-k}(b) we have plotted the minima of $\delta f(v)$ as a function of frequency for different values of the disorder parameters $\Delta m$ and $\Delta k$. As stated above, these functions exhibit a minimum  corresponding to a maximum deviation of the eigenstate fluctuations from the GOE behaviour. We have marked the positions of these minima in the phase diagrams in Fig.\ \ref{fig-phase} and find that they coincide with the values of the BP frequencies. Obviously both the disorder-modified plane waves ($\omega<\omega_{\rm BP}$) as well as the random-matrix states ($\omega>\omega_{\rm BP}$) obey the GOE statistics rather well, whereas the states at the cross-over (i.e. the states with $\omega=\omega_{\rm BP}$) have a maximum deviation from GOE. Of course, approaching the mobility edge the GOE behaviour disappears.

\section{Localization properties of transport states}
\label{sec-loctra}

\subsection{TMM results and the phase diagrams for mass and spring disorder}
\label{sec-TMM_res}

We have performed TMM calculations at $\Delta m, \Delta k= 0.2, 0.4, \ldots, 2$ (see Appendix \ref{sec-tmm} for details). In addition to these disorders, more are required to verify the phase boundary obtained for the pure mass disorder case from direct transformation of the electronic potential disordered phase boundary in section \ref{sec-disorder}. A small selection of additional disorders is chosen as $\Delta m = 2.2, 4, 6, 9$. In the pure spring disordered case a larger additional list is required as a phase boundary is yet to be established. An adequate resolution is achieved with additional disorders of $\Delta k= 2.5, 3, 4, 4.5, 5, 6, 7, 8, 9, 10$. The average of the mass and spring constant disorder ($\overline{m}$ and $\overline{k}$) has been kept fixed at $1$ for all cases. For every disorder value, the reduced localization length, $\Lambda_M$ has been calculated for a range of frequencies and system widths $M= 6, 8, 10$ and $12$ to an accuracy of $0.1\%$  of the variance. 
%
\begin{figure*}[tb]
\centering
\subfigure[\, $n_{r_0}= 2$, $n_{r_1}= 3$, $n_i= 1$, $m_r= 2$, $m_i= 0$]{\label{fig-dm_1.2}\includegraphics[width=0.45\textwidth]{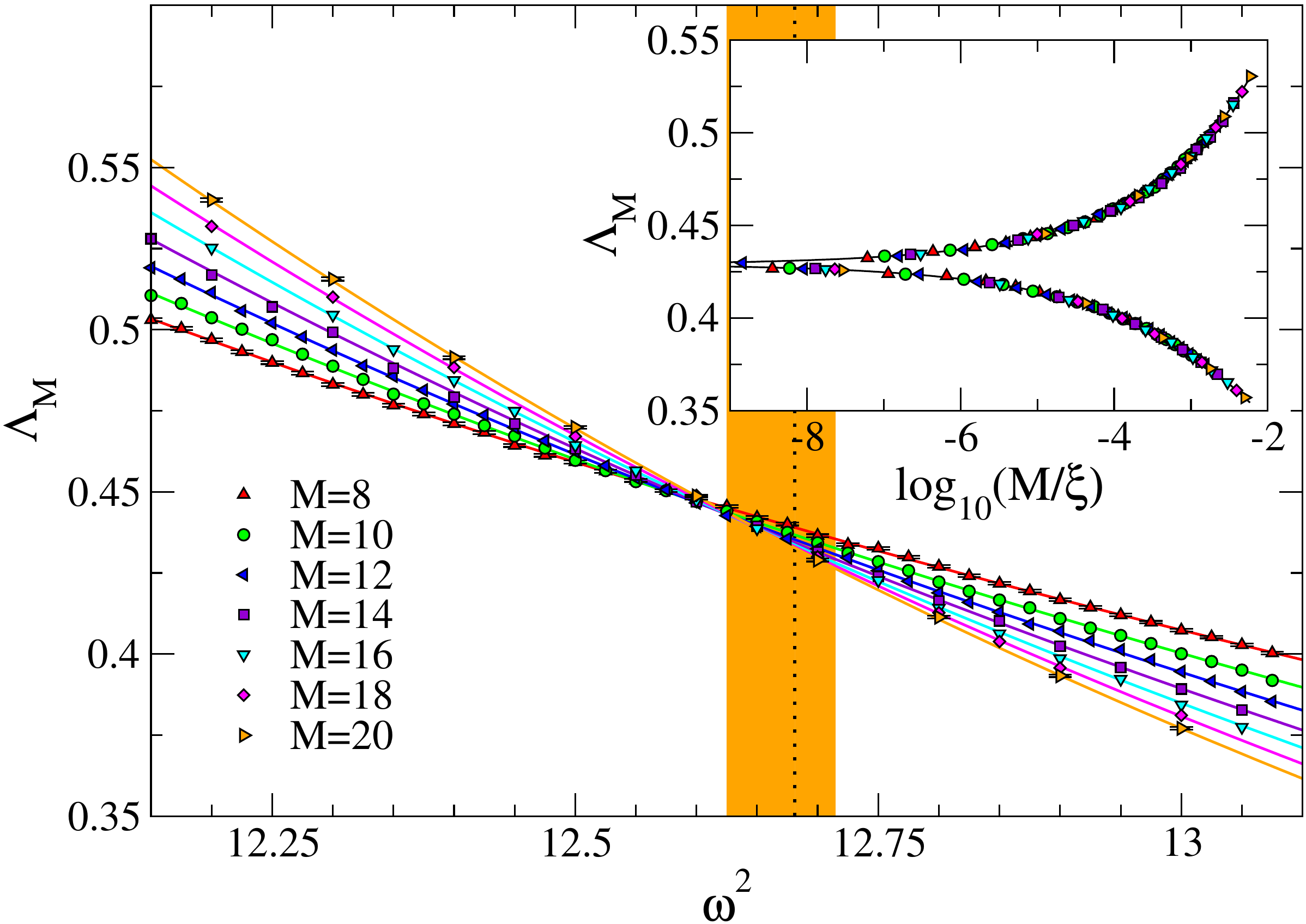}}
%
%
%
%
%
\subfigure[\, $n_{r_0}= 2$, $n_{r_1}= 2$, $n_i= 1$, $m_r= 1$, $m_i= 0$]{\label{fig-dk_7}\includegraphics[width=0.45\textwidth]{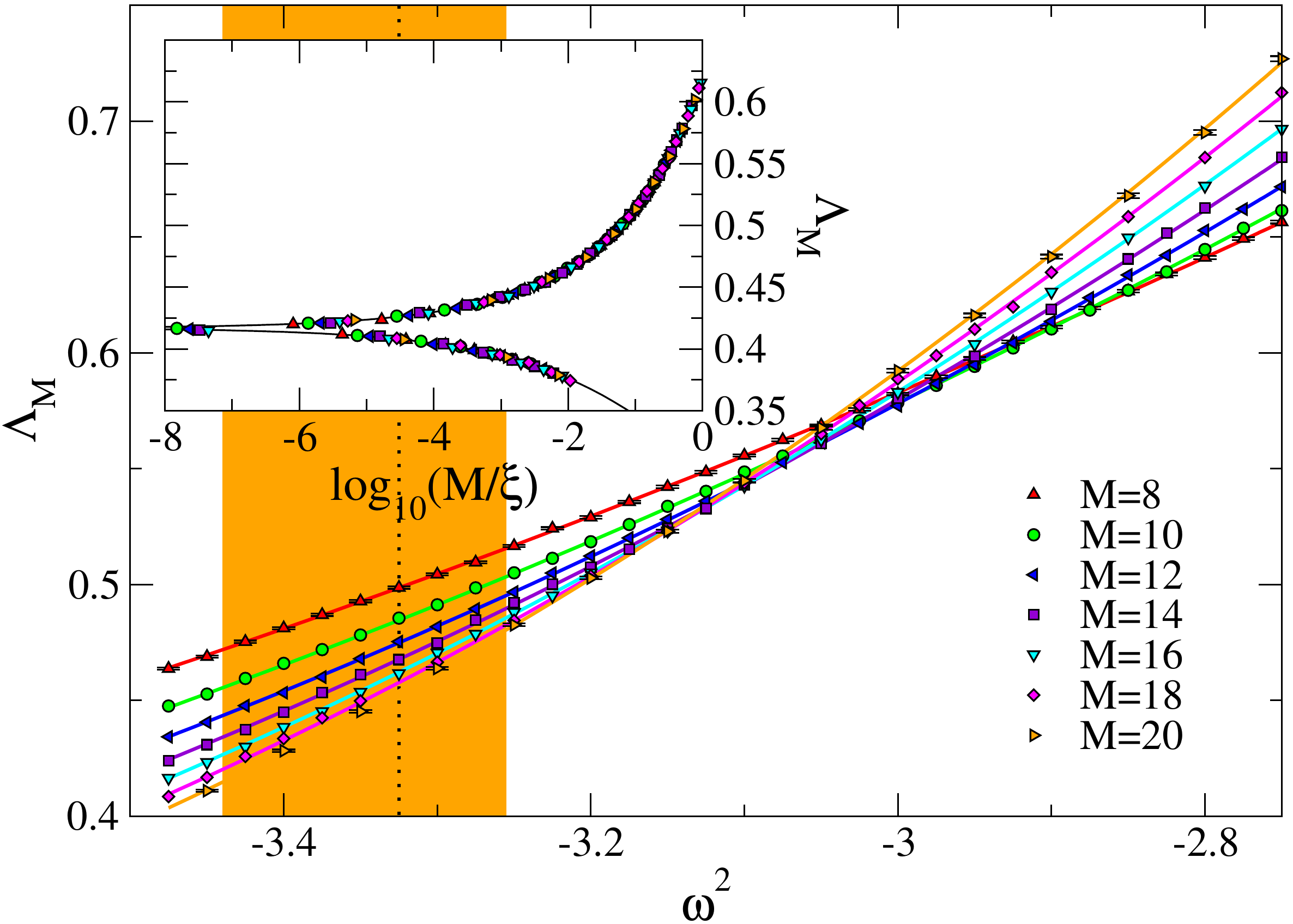}}\\
\caption{(Color online) Reduced localization lengths $\Lambda_M$ plotted as function of $\omega^2$ for various system sizes as indicated by different symbols. Panel (a) shows mass disorder for $\Delta m=1.2$ while (b) is for spring constant disorder of $\Delta k=7$. The lines in each plot show the fits obtained from FSS, the orders of the expansion are given below each figure (see Appendix \ref{sec-fss}). The vertical dotted line represents the estimated values of $\omega_c^2$ with orange shading indicating the error obtained from Monte Carlo analysis (in Tab.\ \ref{tab-criticalparameter-w2}). Error bars are only shown for the largest and smallest system size, as in all cases they are within symbol size. The insets display the obtained scaling function when the irrelevant components have been subtracted.
}
\label{fig-fss-results-omega2}
\end{figure*}

In Figs.\ \ref{fig-dm_1.2}--\ref{fig-dk_7}, we show the resulting disorder and $\omega^2$ dependencies for $2$ of the $6$ representative mass/spring disorder regions. At all disorder magnitudes for both spring constant and mass disorder, these figures reveal clear transitions from extended behaviour, with increasing $\Lambda_M$ values for increasing $M$, to localized behaviour, where $\Lambda_M$ decreases when $M$ increases. We also see in these figures frequency regions where $\Lambda_M$ remains roughly constant upon changing $M$. Such regions are in the vicinity of a change from delocalization to localization and hence Figs.\ \ref{fig-dm_1.2}--\ref{fig-dk_7} indicate the existence of a delocalization-localization transition. We roughly estimate the transition regions by the frequency value at which the values of $\Lambda_M$ for the largest and the second largest system size cross ($M=10,12$). Then we obtain a similarly rough estimate of the error of this estimate from the difference with respect to the frequency value which we obtain when we take the crossing point between the largest and smallest system sizes ($M=6,12$). These estimates are the basis of the phase diagrams in Fig.\ \ref{fig-phase}.

In the spring constant disorder case we need to pay special attention to the ${k^{-1}_{x+1}}$ term in equation \eqref{eq-TMM} as at disorders $\Delta k \geq 2$, the disorder distribution contains values close to zero which when applied in the ${k^{-1}_{x+1}}$ can dramatically increase a single site amplitude dwarfing surrounding amplitudes. We apply a cut-off whereby if $|k_{x+1}| \le 10^{-4} \overline{k}$ the value is rejected and another randomly chosen. We re-estimate all transition frequencies of previously mentioned disorders and find that the new estimates are identical within the previous error bars and therefore keep the estimates obtained with unaltered distributions.

We plot these estimates of the critical frequencies in the phase diagrams of Figs.\ \ref{fig-phase}(a) and \ref{fig-phase}(b). As we can see, for the pure mass disorder case, Fig.\ \ref{fig-phase}(a) very well reproduces the estimated phase diagram obtained from comparison with the electronic phase diagram in the Anderson model.\cite{PinSR12}
Most interestingly, the small pocket of extended states in the complex frequency spectrum of the mass disorder phase diagram is clearly identified by the two transitions from localized to delocalized and back to localized at $\Delta m= 9$. 

For the pure spring constant disorder, we see that in the region $0\leq \omega^2 < 12$, all states remain extended up to the largest considered spring constant disorder $\Delta k=10$. This is similar to the electronic case with pure hopping disorder \cite{CaiRS99,BisCRS00,Cai98T} where even very strong hopping disorder does not lead to complete localization close to $E=0$.\cite{Note2}

We find that both for mass and spring constant disorder, the $\omega^2=0$ mode \cite{LudTE03,LudSTE01,Rus02,ShiNN07} remains extended regardless of the disorder strength. This is in agreement with previous studies in one- and two-dimensional systems.\cite{LudTE03} We also observe for both mass and spring constant disorder very strong shifts of the crossing points of $\Lambda_M$ when changing $M$. This is to be expected since we are effectively dealing with transition regions in the vicinity of the tails of the VDOS (cp.\ Fig.\ \ref{fig-vdos}) and hence the systematic size changes are also strongly influenced by non-universal changes in the VDOS. This is similar to the situation for the electronic case where the transition at the mobility edges for $E\neq 0$ is known to be more difficult to study.\cite{MacK81,KraBMS90}

\subsection{FSS estimates for the critical parameters}
\label{sec-critdis_res}

In order to obtain more reliable estimates for the transition point $\omega^2_c$ as well as to ascertain the existence of a divergent correlation length $\xi(\omega) \propto |\omega^2 - \omega^2_c |^{-\nu}$ at $\omega^2_c$ with critical exponent $\nu$, we need to proceed to the $M\rightarrow\infty$ limit. This we do, as in the electronic case, via an FSS procedure (see Appendix \ref{sec-fss} for details).\cite{SleO99a} 
We perform the FSS analysis on the raw data of reduced localization lengths $\Lambda_M$ as functions of $\omega^2$ as well as $\omega$ (with $\xi(\omega) \propto |\omega - \omega_c |^{-\nu}$). While the latter seems more natural in the context of vibrations, we emphasise that the former is more convenient when comparing to the electronic case where $\omega_c^2$ is related to the energy.\cite{PinSR12}

For both pure mass and pure spring disorder, we concentrate on $3$ disorder values each, choosing those from the $3$ different domains of the phase diagrams of Figs.\ \ref{fig-phase}(a) and \ref{fig-phase}(b), namely (i) $\omega^2\geq 0$; $\Delta m, \Delta k < 2$, (ii) $\omega^2 \geq 0$; $\Delta m, \Delta k \geq 2$ and (iii) $\omega^2<0$. 
For these $6$ points, we compute additional high-precision data for $M= 14, 16, 18$ and $20$. The additional $\Lambda_M$ values for two of these $6$ transitions have also been shown in Figs.\ \ref{fig-dm_1.2} -- \ref{fig-dk_7}. We then apply the FSS procedure of appendix \ref{sec-fss} and hence obtain precise estimates of the critical parameters and transition frequencies $\omega_c^2$ of a vibrating solid in the thermodynamic limit. These $\omega_c^2$ values have also been indicated in the phase diagrams as in Figs.\ \ref{fig-phase}(a) and \ref{fig-phase}(b)). 
In Tab.\ \ref{tab-criticalparameter-w2} we show the results for the high-precision FSS analysis. We find that in all cases, a consistent, robust and stable fit with quality-of-fit parameter $\Gamma_q$ larger than $0.1$ can be identified. In particular, the FSS for $\omega$ as well as $\omega^2$ gives consistent results.
\begin{center}
\begin{table*}[tb]
\begin{tabular}{ccccccccccccccc}
\hline
\hline
$\Delta m$ & $M$ & $\omega$ & $\omega^2$ & $n_{r_0}$ & $n_{r_1}$ & $n_i$ & $m_r$ & $m_i$ & $\omega_c$ & $\omega^2_c$ & $\nu$ & $\chi^2$ & $\mu$ & $\Gamma_q$\\
\hline
1.2 & 8 -- 20 & & [12.15, 13.1] & 2 & 3 & 1 & 2 & 0 & & $12.681^{+0.056}_{-0.034}$ & $1.57^{+0.14}_{-0.09}$ & $165^{+38}_{-34}$ & 165 & 0.84\\
4.0 & 8 -- 20 & & [3.75, 4.25] & 3 & 2 & 1 & 1 & 0 & & $4.134^{+0.024}_{-0.020}$ & $1.57^{+0.06}_{-0.08}$ & $572^{+69}_{-64}$ & 574 & 0.99\\
9.0 & 8 -- 20 & & [-1.65, -1.5] & 2 & 3 & 1 & 2 & 0 & & $-1.623^{+0.018}_{-0.037}$ & $1.56^{+0.41}_{-0.18}$ & $154^{+37}_{-33}$ & 154 & 0.87\\
%
1.2 & 8 -- 20 & [3.485, 3.62] & & 2 & 3 & 1 & 2 & 0 & $3.561^{+0.008}_{-0.005}$ & & $1.57^{+0.15}_{-0.09}$ & $164^{+38}_{-34}$ & 165 & 0.84\\
4.0 & 8 -- 20 & [1.936, 2.062] & & 3 & 2 & 1 & 1 & 0 & $2.033^{+0.006}_{-0.005}$ & & $1.55^{+0.07}_{-0.08}$ & $573^{+65}_{-63}$ & 573 & 0.99\\
9.0 & 8 -- 20 & [-1.284, -1.225] & & 2 & 3 & 1 & 1 & 0 & $-1.273^{+0.006}_{-0.014}$ & & $1.56^{+0.44}_{-0.17}$ & $155^{+36}_{-33}$ & 155 & 0.83\\
\hline
\hline
$\Delta k$ & $M$ & $\omega$ & $\omega^2$ & $n_{r_0}$ & $n_{r_1}$ & $n_i$ & $m_r$ & $m_i$ & $\omega_c$ & $\omega^2_c$ & $\nu$ & $\chi^2$ & $\mu$ & $\Gamma_q$\\
\hline
1.0 & 10 -- 20 & & [12.48, 12.6] & 3 & 1 & 1 & 1 & 1 & & $12.527^{+0.003}_{-0.004}$ & $1.58^{+0.05}_{-0.04}$ & $132^{+34}_{-30}$ & 132 & 0.62\\
10.0 & 6 -- 16 & & [18.8, 20.3] & 1 & 3 & 1 & 2 & 0 & & $19.749^{+0.043}_{-0.038}$ & $1.51^{+0.08}_{-0.08}$ & $176^{+39}_{-36}$ & 176 & 0.84\\
7.0 & 8 -- 20 & & [-3.5, -2.75] & 2 & 2 & 1 & 1 & 0 & & $-3.325^{+0.070}_{-0.115}$ & $1.59^{+0.23}_{-0.29}$ & $162^{+38}_{-33}$ & 162 & 0.51\\
1.0 & 10 -- 20 & [3.529, 3.55] & & 3 & 3 & 1 & 1 & 2 & $3.540^{+0.001}_{-0.001}$ & & $1.47^{+0.15}_{-0.05}$ & $157^{+39}_{-34}$ & 156 & 0.49\\
10.0 & 6 -- 16 & [4.335, 4.506] & & 2 & 3 & 1 & 2 & 0 & $4.441^{+0.008}_{-0.009}$ & & $1.52^{+0.15}_{-0.53}$ & $199^{+41}_{-38}$ & 199 & 0.87\\
7.0 & 8 -- 20 & [-1.87, -1.66] & & 2 & 2 & 1 & 1 & 0 & $-1.825^{+0.019}_{-0.033}$ & & $1.60^{+0.21}_{-0.19}$ & $162^{+38}_{-34}$ & 162 & 0.79\\
\hline
\end{tabular}
\caption{Values of critical parameters $\omega_\text{c}$, $\omega_\text{c}^2$ and $\nu$ for pure mass (top) and pure spring constant (bottom) disorder computed from FSS performed in the given $M$ and $\omega$, $\omega^2$ ranges and with the orders of the expansion \eqref{eq-fss} given by $n_{r_0}$, $n_{r_1}$, $n_i$, $m_r$ and $m_i$. The minimised $\chi^2$ value, the degrees of freedom $\mu$ and the resulting goodness-of-fit parameter $\Gamma_q$ are also shown for each fit. The errors correspond to non-symmetric $95\%$ confidence intervals (see Appendix \ref{sec-fss}).}
\label{tab-criticalparameter-w2}
\end{table*}
\end{center}

A weighted average of the critical exponent for the estimates in Tab\ \ref{tab-criticalparameter-w2} is $\nu= 1.550^{+0.020}_{-0.017}$. This is in excellent agreement with previous numerical studies of the Anderson model for electron localization which have found the critical exponent $\nu \equiv 1.57 \pm 0.02$.\cite{Mac94,SleO99a,RodVSR10} In the vibrational model, no previous high-precision results are available. With an accuracy of 2\% in the raw TMM data for spring disorder $\Delta k = 1.8$, Akita and Ohtsuki\cite{AkiO98} previously found a critical exponent of $\nu \approx 1.2 \pm 0.2$. Recently, Monthus and Garel\cite{MonG10} assumed $\nu = 1.57$ and showed that their participation ratio data for high disorder collapsed onto a scaling function. All these results for model \eqref{eq-dynmat} are therefore consistent with the orthogonal universality class of the Anderson model.\cite{EveM08}

\section{Conclusions}
\label{sec-concl}


In the preceding sections, we have established the existence and universality of the localization-delocalization transitions for vibrational excitations in a simple harmonic solid at various values of frequency and mass or spring constant disorder. While the model itself is simple, the resulting phase diagrams are not and exhibit intriguing features. In particular, there are regions of localized and extended unstable modes with transitions between them that belong to the same universality class as in the stable regimes. Namely, the universality class of the 3D electronic Anderson metal-insulator transition.\cite{KraM93} Our results show that the FSS scaling works both when using the $\omega$ scaling, most natural from a vibrational point of view, as well as the $\omega^2$ scaling, motivated by the electronic analogue.
The peak in the VDOS as shown in Fig.\ \ref{fig-vdos} seems identifiable as a continuation of the van Hove singularity at low --- mass or spring constant --- disorder. The peak is not visible in the participation ratio data, but its signature can be seen again in the wave function statistics. Whether it can truly be called a boson peak, although it does of course appears as such in $g(\omega)/\omega^2$ plots, remains undetermined at present.\cite{LudTE03}
The wave function statistics of section \ref{sec-PTS} and the plots of critical vibrational amplitudes in Fig.\ \ref{fig-Mstates} and \ref{fig-Kstates} also reveal subtle differences between mass and spring disorder. A more in-depth analysis of the multifractal properties and scaling properties of the generalised participation ratio at the transition might be very useful. However, we note that previous studies in fluids\cite{HuaW10} and elastic beads\cite{FaeSPL09} have found good agreement with the multifractal spectrum obtained for the electronic case.\cite{VasRR08,RodVR08} 

Making contact with possible experimental systems, we note that the transitions are at rather high frequencies. The Debye temperatures $\Theta_D=\hbar \omega_D/k_B$ of, e.g., Si and Ge --- candidate materials for milli-Kelvin cooling devices\cite{ClaMWR05,ZebEDR12} whose study got us interested in this research --- are $\Theta_D= 645$K and $374$K, respectively. Assuming that the upper band edge of the clean case can be approximated by the respective Debye frequencies $\omega_D = 1.34 \times 10^{13}$Hz and $7.79 \times 10^{12}$Hz, respectively, we see from the phase diagrams that the transition frequencies remain quite high. Localization of vibrations for these systems in the stable regime  appears only possible for frequencies in or above the far infrared frequency spectrum, particularly for spring constant disorder. The transition for very large mass disorder does tend towards smaller $\omega^2$ values, but these mass disorders are already deep in the unstable regime $\Delta m>2$. This is of course dramatically different from the electronic situation where a disorder of $16.55$ is known to localize all states in a simple cubic system with band width $12$ (in units of hopping strength).\cite{KraM93}
We note that the unstable regions of the phase diagrams for $\Delta m$, $\Delta k>2$ with possibly negative masses and spring constants are now recognised to be of considerable interest for acoustic and disordered metamaterial applications. \cite{LiuZMZ00,Hir04,ChaLF06,FanXXA06,DinLQS07,YaoZH08,ZhaYF09,HuaS09,WriC09,HeQCX10} Here our identification of regions of extended states should prove useful.

\acknowledgments
We gratefully acknowledge discussions with Evan Parker and Alberto Rodriguez-Gonzalez as well as the EPSRC (EP-F040784-1) and the EC ``Nanofunction'' network of excellence for financial support.

\appendix

\section{Coherent potential approximation}
\label{sec-cpa}

As an estimate of the VDOS calculations of section \ref{sec-vdos} we compute the VDOS using the coherent potential approximation (CPA).\cite{YonM73} In the spring constant disorder case we introduce a frequency dependant force constant (``self energy'') $\Gamma(z)$ and determine its contribution self-consistently using the scattering matrix formalism,\cite{SchDG98}
\begin{equation}
\left\langle \frac{\Gamma(z) - k_{ij}}{1-\frac{[\Gamma(z) - k_{ij}][1-zG(z)]}{3\Gamma(z)}}\right\rangle=0,
\end{equation}
where $z=\omega^2+i0_+$ is the regularised complex frequency. The local Green function of the effective medium is
\begin{equation}
G(z)=\frac{1}{\Gamma(z)}G_0\left(\frac{z}{\Gamma(z)}\right)
\end{equation}
and $G_0$ is the Green function  for the clean system.\cite{Joy73} The averaged VDOS is then given as
\begin{equation}
\langle g(\omega^2)\rangle=-\frac{1}{\pi}\Im[G(z)].
\end{equation}

In the mass disordered case we use the transformation rule (\ref{transformation}) to map the problem to an Anderson problem with fluctuating local
energies $\epsilon_i$ and then use the conventional single-site
CPA.\cite{Yon68} The self energy $\Sigma(z)$ with $z=E+i0^+$ is
given by setting the following CPA scattering matrix equal to zero:
\begin{equation}
\langle t\rangle=\left\langle \frac{\epsilon_{i_0} - \Sigma(z)}{1-(\epsilon_{i_0} - \Sigma(z)) G_0[z-\Sigma(z)]} \right\rangle = 0
\label{equ-CPA-avgt}
\end{equation}
The single-site CPA problem is known to exhibit rather unstable iteration properties. We obtained a good iteration performance using the following iteration method \cite{WolM02}, which is equivalent to the CPA condition (\ref{equ-CPA-avgt}).
\begin{equation}
\Sigma^{(n+1)}(z)=\Sigma^{(n)}(z)+\frac{\langle t\rangle^{(n)}}{1+\langle t\rangle^{(n)}G_0[z-\Sigma^{(n)}(z)]},
\end{equation}
\begin{equation}
\langle t\rangle^{(n)}=  \left\langle \frac{\epsilon_{i_0} - \Sigma^{(n)}(z)}{1-[\epsilon_{i_0} - \Sigma^{(n)}(z)] G_0[z-\Sigma^{(n)}(z)]} \right\rangle,
\end{equation}
where $n$ is the iteration count.
The average density of states is then calculated from the Green's function as
\begin{equation}
\langle g(E)\rangle = -\frac{1}{\pi} \Im \left\{ G_0\left[z-\Sigma(z)\right]\right\} .
\end{equation}
%
The results for both disorders are shown in Fig.\ \ref{fig-vdos} as thin dashed lines next to the numerical VDOS. We find good agreement between CPA results and the numerical calculations for both weak and strong disorder and in the stable ($\omega^2>0$) and unstable ($\omega^2<0$) spectral regions.

\section{The transfer-matrix approach}
\label{sec-tmm}

The transfer-matrix method (TMM) allows for a very memory efficient way to iteratively calculate the decay length $\Lambda_M$ of vibrations in a quasi-one dimensional bar with cross section $M\times M$ for lengths $L \gg M$. Equation (\ref{eq-dyna}) has to be rearranged into a form where the amplitude of vibration of a site in layer $x+1$ --- when $x$ is chosen as the direction of transfer --- is calculated solely from parameters of sites in previous layers $x$ and $x-1$,
\begin{eqnarray}
 u_{x+1,y,z} &=& -\frac{1}{k_{x+1,y,z}} \left[ (\omega^2 m_{x,y,z} + k_{\text{all}})u_{x,y,z} -h_{x} \right] \nonumber\\
  && \mbox{~~} - \frac{k_{x-1,y,z}}{k_{x+1,y,z}} u_{x-1,y,z} \label{eq-TMM_singles}
\end{eqnarray}
Here $h_{x} \equiv k_{x,y,z+1} u_{x,y,z+1} + k_{x,y,z-1} u_{x,y,z-1} + k_{x,y+1,z} u_{x,y+1,z} + k_{x,y-1,z} u_{x,y-1,z}$ denotes the collection of in-plane contributions to the final amplitude, $k_{\text{all}}=k_{x,y,z+1} + k_{x,y,z-1} + k_{x,y+1,z} + k_{x,y-1,z} + k_{x+1,y,z} + k_{x-1,y,z}$ and we have changed back to the explicit notation such that $u_j\equiv u_{x,y,z}$ for $\vec{r}_j=(x,y,z)_j$. Similarly, $k_{jl}\equiv k_{x+1,y,z}$ for $\vec{r}_l=(x+1,y,z)_l$.
With $U_x =\left(u_{x,1,1}, u_{x,1,2}, u_{x,2,1}, \dots, u_{x,M,M} \right)$, we can define $U_x$, $U_{x+1}$ and $U_{x-1}$ as vectors containing the amplitudes of the constituent sites in layers $x$, $x+1$ and $x-1$, respectively. Equation \eqref{eq-TMM_singles} can now be expressed in standard transfer-matrix form 
\begin{equation}
 \left[ \begin{array}{c}
   U_{x+1} \\ U_x
 \end{array} \right] =
\underbrace{
 \left[ \begin{array}{cc}
   -\frac{ \left[ \left(\omega^2 m_x + k_{\text{all}}\right) \mathbf{1} -\mathbf{H}_x \right] }{k_{x+1}} & -\frac{k_{x-1}}{k_{x+1}} \mathbf{1} \\ \mathbf{1} &
\mathbf{0}
 \end{array} \right]}_{\mathbb{T}_x}
 \left[ \begin{array}{c}
   U_x \\ U_{x-1}
 \end{array} \right] \label{eq-TMM},
\end{equation}
where $\mathbf{H}_{x}$ is a $M\times M$ matrix containing all in-layer contributions, $\mathbf{0}$ and $\mathbf{1}$ are the zero and unit matrices, respectively. 

Formally, the transfer matrix $\mathbb{T}_x$ is used to `transfer' vibrational amplitudes $U$ from one slice to the next and repeated multiplication of this gives the global transfer matrix $\tau_L = \prod^L_{x=1} \mathbb{T}_x$. The limiting matrix $\Gamma \equiv \lim_{L\to\infty} \left(\tau_L \tau_L^\dagger \right)^{\frac{1}{2L}}$ exists\cite{Ose68} and has eigenvalues $e^{\pm \gamma_i}$, $i=1, \ldots, M$. The inverse of these Lyapunov exponents $\gamma_i$ are estimates of decay/localization lengths and the physically relevant largest vibrational decay length is $\lambda_M(\omega^2) =1/\text{min}_i\left[\gamma_i(\omega^2)\right]$. The reduced (dimensionless) decay length may then be calculated as $\Lambda_M(\omega^2) =\lambda_M(\omega^2)/M$.

\section{Finite-size scaling}
\label{sec-fss}

The FSS includes two types of corrections to scaling, namely, those which account for the nonlinearities of the $\Delta m$, $\Delta k$ dependence of the scaling variables (relevant scaling) and for the mentioned shift of the point at which the $\Lambda_M(\omega^2)$ curves cross (irrelevant scaling). The starting point for the FSS in terms of $\omega^2$ is the scaling ansatz
\begin{equation}
 \Lambda_M (\omega^2) = f\left( \chi_r M^{\frac{1}{\nu}}, \chi_i M^y \right),
\label{eq-fss}
\end{equation}
where $\chi_r$ and $\chi_i$ are the relevant and irrelevant scaling variables, respectively. The function $\Lambda_M$ is then Taylor expanded up to
the order $n_i$ and we have
$
 \Lambda_M = \sum^{n_i}_{n=0} \chi_i^n M^{ny} f_n \left(\chi_r M^{\frac{1}{\nu}} \right)
$ 
from where we obtain a series of functions $f_n$ which are in turn Taylor expanded up to an order $n_r$ such that
$
 f_n \left( \chi_r M^{\frac{1}{\nu}} \right) = \sum^{n_r}_{k=0} a_{nk}\chi_r^k M^{\frac{k}{\nu}}
$. 
Nonlinearities are taken into account by expanding both $\chi_i$ and $\chi_r$ in terms of the dimensionless frequency $w=(\omega^2_\text{c} -\omega^2)/
\omega^2_\text{c}$ such that
$
 \chi_r(w) = \sum^{m_r}_{m=1} b_m w^m, \chi_i(w) = \sum^{m_i}_{m=0} c_m w^m
$ 
where the orders of the expansions are $m_r$ and $m_i$. For a more rigorous analysis we hard-code the zero-th and first order of the irrelevant expansion and Taylor expand each appearance of $f_n$ separately.\cite{RodVSR10} 

The expansions of the fit functions and the fit are performed numerically up to the orders $n_{r_0}$, $n_{r_1}$, $n_i$, $m_i$ and $m_r$. Each individual data set can be best suited to a particular expansion, the general rule being that the orders of expansion should be kept as low as possible while giving the best fit to the data, and minimising the estimated standard errors for the critical parameters $\omega_c^2$ and $\nu$. We check for stability of the fit by individually increasing each expansion parameter by one and checking to see that the obtained parameters remain within the $95\%$ confidence intervals of the original fit.

The confidence intervals are then recomputed through a Monte Carlo analysis.\cite{RodVSR10} We obtain a perfect data series from the fit with the previously computed expansion. We next vary each data point according to a Gaussian distribution with the right standard deviation. With this synthetic data, we then repeat the FSS fit to obtain new estimates of the critical parameter. We repeat this operation $5000$ times and compute the distribution function for each critical parameter. We then estimate the true errors from these histograms by taking as errors those values at which $2.5\%$ of the distribution are below or above bulk.

%
%

\begin{thebibliography}{10}

\bibitem{And58}
P.~W. Anderson, Phys. Rev. {\bf 109},  1492  (1958).

\bibitem{LeeR85}
P.~A. Lee and T.~V. Ramakrishnan, Rev. Mod. Phys. {\bf 57},  287  (1985).

\bibitem{KraM93}
B. Kramer and A. MacKinnon, Rep. Prog. Phys. {\bf 56},  1469  (1993).

\bibitem{BelK94}
D. Belitz and T.~R. Kirkpatrick, Rev. Mod. Phys. {\bf 66},  261  (1994).

\bibitem{EveM08}
F. Evers and A.~D. Mirlin, Rev. Mod. Phys. {\bf 80},  1355  (2008).

\bibitem{BilJZB08}
J. Billy {\it et~al.}, Nature {\bf 453},  891  (2008).

\bibitem{RoaDFF08}
G. Roati {\it et~al.}, Nature {\bf 453},  895  (2008).

\bibitem{FaeSPL09}
S. Faez, A. Strybulevych, J.~H. Page, A. Lagendijk and B.~A. vanTiggelen, Phys. Rev. Lett. {\bf 103},  155703  (2009).

\bibitem{SchDG98}
W. Schirmacher, G. Diezemann, and C. Ganter, Phys. Rev. Lett. {\bf 81},  136
  (1998).

\bibitem{KanRB01}
J.~W. Kantelhardt, S. Russ, and A. Bunde, Phys. Rev. B {\bf 63},  064302
  (2001).

\bibitem{PinSR12}
S. Pinski, W. Schirmacher, and R. {R\"{o}mer}, Europhys. Lett. {\bf 97},  16007
   (2012).

\bibitem{Rus02}
S. Russ, Phys. Rev. B {\bf 66},  012204  (2002).

\bibitem{SchD99}
W. Schirmacher and G. Diezemann, Ann. Phys. (Leipzig) {\bf 8},  727  (1999).

\bibitem{FelKAW93}
J.~L. Feldman, M.~D. Kluge, P.~B. Allen, and F. Wooten, Phys. Rev. B {\bf 48},  12589
  (1993).

\bibitem{SarMP04}
S.~K. Sarkar, G.~S. Matharoo, and A. Pandey, Phys. Rev. Lett. {\bf 92},  215503
   (2004).

\bibitem{CanV85}
J. Canisius and J. {van Hemmen}, J. Phys. C {\bf 18},  4873  (1985).

\bibitem{LudSTE01}
J. Ludlam, T. Stadelmann, S. Taraskin, and S. Elliott, Journal of
  Non-Crystalline Solids {\bf 293},  676  (2001).

\bibitem{ShiNN07}
H. Shima, S. Nishino, and T. Nakayama, J. Phys.: Conf. Ser. {\bf 92},  012156
  (2007).

\bibitem{LudTE03}
J.~J. Ludlam, S.~N. Taraskin, and S.~R. Elliott, Phys. Rev. B {\bf 67},  132203  (2003).

\bibitem{BemL95}
S.~D. Bembenek and B.~B. Laird, Phys. Rev. Lett. {\bf 74},  936  (1995).

\bibitem{BemL96}
S.~D. Bembenek and B.~B. Laird, J. Chem. Phys. {\bf 104},  5199  (1996).

\bibitem{HuaW10}
B.~J. Huang and T.-M. Wu, Phys. Rev. E {\bf 82},  051133  (2010).

\bibitem{HuaW09}
B.~J. Huang and T.-M. Wu, Phys. Rev. E {\bf 79},  041105  (2009).

\bibitem{LiuZMZ00}
Z. Liu {\it et~al.}, Science {\bf 289},  1734  (2000).

\bibitem{Hir04}
M. Hirsekorn, Appl. Phys. Lett. {\bf 84},  3364  (2004).

\bibitem{ChaLF06}
C. Chan, J. Li, and K. Fung, Science A {\bf 7},  24  (2006).

\bibitem{ZhaYF09}
S. Zhang, L. Yin, and N. Fang, Phys. Rev. Lett. {\bf 102},  194301  (2009).

\bibitem{DinLQS07}
Y. Ding, Z. Liu, C. Qiu, and J. Shi, Phys. Rev. Lett. {\bf 99},  093904
  (2007).

\bibitem{WriC09}
D.~W. Wright and R.~S. Cobbold, Ultrasound {\bf 17},  68  (2009).

\bibitem{BulSK87}
B. Bulka, M. Schreiber, and B. Kramer, Z. Phys. B {\bf 66},  21  (1987).

\bibitem{AkiO98}
Y. Akita and T. Ohtsuki, J. Phys. Soc. Jap. {\bf 67},  2954  (1998).

\bibitem{BorH54}
M. Born and K. Huang, {\em Dynamical Theory of Crystal Lattices} (Oxford, Univ.
  Press, New York, 1954).

\bibitem{Sri90}
G. Srivastava, {\em The Physics of Phonons} (Taylor \& Francis Group, 270
  Madison Avenue, New York, 1990).

\bibitem{BraK03}
{\em The {Anderson} Transition and its Ramifications --- Localisation, Quantum
  Interference, and Interactions}, Vol.~630 of {\em Lecture Notes in Physics},
  edited by T. Brandes and S. Kettemann (Springer, Berlin, 2003).

\bibitem{GruS95}
H. Grussbach and M. Schreiber, Phys. Rev. B {\bf 51},  663  (1995).

\bibitem{MadK93}
B. Madan and T. Keyes, Journal of Chemical Physics {\bf 98},  4  (1993).

\bibitem{PinR11}
S.~D. Pinski and R.~A. {R\"{o}mer}, J.\ Phys.:\ Conf.\ Ser. {\bf 286},  012025
  (2011).

\bibitem{LehSY98}
R.~B. Lehoucq, D.~C. Sorensen, and C. Yang, {\em Arpack User's Guide: Solution
  of Large-Scale Eigenvalue Problems With Implicityly Restorted Arnoldi
  Methods} (Society for Industrial Mathematics, Philaqdelphia, PA., 1998).

\bibitem{SchBR06}
O. Schenk, M. Bollh\"{o}fer, and R. R\"{o}mer, SIAM Journal of Sci. Comp. {\bf
  28},  963  (2006).

\bibitem{Note1}
For the symmetric case alone, the {\protect \sc Jadamilu} package is also very
  useful.\protect \cite {SchBR06}.

\bibitem{VasRR08}
L.~J. Vasquez, A. Rodriguez, and R.~A. R\"{o}mer, Phys. Rev. B {\bf 78},
  195106  (2008), arXiv: cond-mat:0807.2217v1.

\bibitem{LudTED05}
J.~J. Ludlam, S.~N. Taraskin, S.~R. Elliot, and D.~A. Drabold, J. Phys.:
  Condens. Matter {\bf 17},  L321  (2005).

\bibitem{MirFME06}
A.~D. Mirlin, Y.~V. Fyodorov, A. Mildenberger, and F. Evers, Phys. Rev. Lett.
  {\bf 97},  046803  (2006).

\bibitem{AndBBB87}
E. Anderson {\it et~al.}, {\em LAPACK Users' Guide} (Society for Industrial
  Mathematics, Philaqdelphia, PA., 1987).

\bibitem{LeoTWB06}
F. L\'{e}onforte, A. Tanguy, J.~P. Wittmer, and J.-L. Barrat, Phys. Rev. Lett.
  {\bf 97},  055501  (2006).

\bibitem{LeoBTW05}
F. L\'{e}onforte, R. Boissiere, A. Tanguy, J.~P. Wittmer, and J.-L. Barrat, Phys. Rev. B {\bf 72},  224206  (2005).

\bibitem{MonM09}
G. Monaco and S. Mossa, PNAS {\bf 106},  16907  (2009).

\bibitem{YonM73}
F. Yonezawa and K. Morigaki, Prog. Theor. Phys. Supplement {\bf 53},  1
  (1973).

\bibitem{ScoSAA06}
T. Scopigno, J.~B. Suck, R. Angelini, F. Albergamo and G. Ruocco, Phys. Rev. Lett. {\bf 96},  135501  (2006).

\bibitem{TarLNE02}
S. Taraskin, J. Ludlam, G. Natarajan, and S. Elliott, Philosophical Magazine B
  {\bf 82},  197  (2002).

\bibitem{MonG10}
C. Monthus and T. Garel, Phys. Rev. B {\bf 81},  224208  (2010).

\bibitem{Meh91}
M.~L. Mehta, {\em Random Matrices and the Statistical Theory of Energy levels}
  (Academic Press, New York, 1991).

\bibitem{Haa92}
F. Haake, {\em Quantum Signatures of Chaos}, 2nd ed. (Springer, Berlin, 1992).

\bibitem{Por65}
C.~E. Porter, {\em Statistical Theories of Spectra: Fluctuations} (Academic
  Press, New York, 1965).

\bibitem{Dys62}
F.~J. Dyson, J. Math. Phys. {\bf 3},  140  (1962).

\bibitem{FyoM94}
Y.~V. Fyodorov and A.~D. Mirlin, Int. J. Mod. Phys. B {\bf 8},  3795  (1994).

\bibitem{UskMRS00}
V. Uski, B. Mehlig, R.~A. {R\"{o}mer}, and M. Schreiber, Phys. Rev. B {\bf 62},
   R7699  (2000).

\bibitem{Nik01b}
B.~K. Nikoli\'{c}, Phys. Rev. B {\bf 64},  014203  (2001).

\bibitem{MulMMS97}
K. M\"uller, B. Mehlig, F. Milde, and M. Schreiber, Phys. Rev. Lett. {\bf 78},
  215  (1997).

\bibitem{CaiRS99}
P. Cain, R.~A. R\"{o}mer, and M. Schreiber, {Ann. Phys. (Leipzig)} {\bf 8},
  SI33  (1999), arXiv: cond-mat/9908255.

\bibitem{BisCRS00}
P. Biswas, P. Cain, R.~A. R\"{o}mer, and M. Schreiber, phys. stat. sol. (b)
  {\bf 218},  205  (2000), arXiv: cond-mat/0001315.

\bibitem{Cai98T}
P. Cain, Master's thesis, Technische {Universit\"{a}t} Chemnitz, 1998.

\bibitem{Note2}
The $E=0$ states are special in the chirally symmetric hopping disorder case,
  whereas we are not aware of any such circumstance in the present case of pure
  spring disorder.

\bibitem{MacK81}
A. MacKinnon and B. Kramer, Phys. Rev. Lett. {\bf 47},  1546  (1981).

\bibitem{KraBMS90}
B. Kramer, A. Broderix, A. MacKinnon, and M. Schreiber, Physica A {\bf 167},
  163  (1990).

\bibitem{SleO99a}
K. Slevin and T. Ohtsuki, Phys. Rev. Lett. {\bf 82},  382  (1999), arXiv:
  cond-mat/9812065.

\bibitem{Mac94}
A. MacKinnon, J. Phys.: Condens. Matter {\bf 6},  2511  (1994).

\bibitem{RodVSR10}
A. Rodriguez, L.~J. Vasquez, K. Slevin, and R.~A. R\"omer, Phys. Rev. Lett.
  {\bf 105},  046403  (2010).

\bibitem{RodVR08}
A. Rodriguez, L.~J. Vasquez, and R.~A. R{\"o}mer, Phys. Rev. B {\bf 78},
  195107  (2008), cond-mat:0807.2209v1.

\bibitem{ClaMWR05}
A. Clark {\it et~al.}, Appl. Phys. Lett. {\bf 86},  173508  (2005).

\bibitem{ZebEDR12}
M. Zebarjadi {\it et~al.}, Energy Environ. Sci. {\bf 5},  5147  (2012).

\bibitem{FanXXA06}
N. Fang {\it et~al.}, Nature Materials {\bf 5},  452  (2006).

\bibitem{YaoZH08}
S. Yao, X. Zhou, and G. Hu, New J. of Phys. {\bf 10},  043020  (2008).

\bibitem{HuaS09}
H. Huang and C. Sun, New J. of Phys. {\bf 11},  013003  (2009).

\bibitem{HeQCX10}
Z. He {\it et~al.}, Europhys. Lett. {\bf 91},  54004  (2010).

\bibitem{Joy73}
G.~S. Joyce, Phil. Trans. R. Soc. A {\bf 273},  583  (1973).

\bibitem{Yon68}
F. Yonezawa, Prog. Theor. Phys. {\bf 40},  734  (1968).

\bibitem{WolM02}
M. Wo\l{}oszyn and A.~Z. Maksymowicz, TASK quarterly {\bf 6},  4  (2002).

\bibitem{Ose68}
V.~I. Oseledec, Trans. Moscow Math. Soc. {\bf 19},  197  (1968).

\end{thebibliography}

\end{document}